\documentclass[a4paper,11pt]{article}
\pdfoutput=1 
\usepackage{jheppub} 
\usepackage{amsmath,amsfonts,amssymb,mathrsfs,graphicx,color,longtable,bm}
\usepackage{hyperref}
\usepackage{graphicx}
\usepackage{array}
\usepackage{color}
\usepackage{enumitem}
\usepackage[usenames,dvipsnames,table]{xcolor}
\usepackage[T1]{fontenc} 
\usepackage[utf8]{inputenc}
\usepackage{amsmath,amssymb,epsfig,txfonts,relsize,xspace}

\allowdisplaybreaks

\def\etmiss{\ensuremath{E_{T}^{\mathrm{miss}}}\xspace}

\newcommand{\mttwo}{\ensuremath{m_{T2}}\xspace}

\newcommand{\ttbar}{\ensuremath{t\bar{t}}\xspace}

\newcommand{\btagged}{\ensuremath{b}-tagged\xspace}

\def\TeV{\ifmmode {\mathrm{\ Te\kern -0.1em V}}\else
                   \textrm{Te\kern -0.1em V}\fi}
\def\GeV{\ifmmode {\mathrm{\ Ge\kern -0.1em V}}\else
                   \textrm{Ge\kern -0.1em V}\fi}
 
\definecolor{nicered}{rgb}{0.7,0.1,0.1}
\definecolor{nicegreen}{rgb}{0.1,0.5,0.1}
\definecolor{niceblue}{rgb}{0.0,0.1,0.7}
\hypersetup{colorlinks,citecolor=nicegreen,linkcolor=nicered,urlcolor=niceblue}
                                      
\def\bm#1{\mbox{\boldmath$#1$\unboldmath}}
\def \beq{\begin{equation}}
\def \eeq{\end{equation}}
\def \bea{\begin{eqnarray}}
\def \eea{\end{eqnarray}}

\begin{document}

\preprint{MPP-2021-115}

\title{Searching for pseudo Nambu-Goldstone boson dark matter  production in association with top quarks}

\author[1]{Ulrich Haisch,}

\author[2]{Giacomo Polesello}

\author[1,3]{and Stefan Schulte}

\affiliation[1]{Max Planck Institute for Physics, F{\"o}hringer Ring 6,  80805 M{\"u}nchen, Germany}

\affiliation[2]{INFN, Sezione di Pavia, Via Bassi 6, 27100 Pavia, Italy}

\affiliation[3]{Technische Universit{\"a}t M{\"u}nchen, Physik-Department, 85748 Garching, Germany}

\emailAdd{haisch@mpp.mpg.de}
\emailAdd{giacomo.polesello@cern.ch}
\emailAdd{sschulte@mpp.mpg.de}

\abstract{Pseudo Nambu-Goldstone bosons~(pNGBs) are attractive dark matter~(DM) candidates, since they couple to the Standard Model (SM) predominantly through derivative interactions. Thereby they naturally evade the strong existing limits inferred from DM direct detection experiments. Working in an effective field theory that includes both derivative and non-derivative DM-SM operators, we  perform a detailed phenomenological study of the Large~Hadron~Collider reach for pNGB~DM  production in association with top quarks. Drawing on motivated benchmark scenarios as examples, we compare our results to other collider limits as well as the constraints imposed by DM~(in)direct detection experiments and the relic abundance.  We furthermore explore implications on the viable parameter space of pNGB~DM. In particular, we demonstrate that DM direct detection experiments become sensitive to many pNGB~DM realisations once  loop-induced interactions are taken into account. The~search strategies and pNGB~DM benchmark models that we discuss can serve as a starting point for dedicated experimental analyses by the ATLAS and the~CMS collaborations.   }

\maketitle
\flushbottom

\section{Introduction}
\label{sec:introduction}

Weakly interacting massive particles~(WIMPs) have been the prime dark matter~(DM) candidate for more than three decades because they can give rise to the correct abundance of DM today via thermal freeze-out production. However, the null results from DM direct  and indirect  detection  experiments~(see for instance~\cite{Klasen:2015uma,Schumann:2019eaa}) along with the failure to observe  anomalous missing transverse energy~($\etmiss$) production at the Large~Hadron~Collider~(LHC)~(see~\cite{Aaboud:2019yqu} for an experimental status report) have by now ruled out large portions of the parameter space of the simplest WIMP hypotheses such as the neutralino in supersymmetric theories. 

Compelling  examples of still viable WIMP models are provided by scenarios in which  DM  consists of composite pseudo Nambu-Goldstone bosons~(pNGBs).  Models of this type  can address simultaneously the electroweak~(EW) hierarchy problem of the Standard Model~(SM) and the DM~puzzle~\cite{Frigerio:2012uc}, and as a result have received notable attention in recent years~\cite{Barger:2008jx,Barger:2010yn,Chala:2012af,Marzocca:2014msa,Barnard:2014tla,Fonseca:2015gva,Brivio:2015kia,Kim:2016jbz,Chala:2016ykx,Barducci:2016fue,Ma:2017vzm,Balkin:2017aep,Balkin:2017yns,Gross:2017dan,Alanne:2018xli,Balkin:2018tma,Ishiwata:2018sdi,Huitu:2018gbc,Karamitros:2019ewv,Davoli:2019tpx,Ruhdorfer:2019utl,Ramos:2019qqa,Arina:2019tib,Abe:2020iph,Okada:2020zxo,Xing:2020uaf,Okada:2021qmi,Coito:2021fgo}. In models in which both the SM Higgs boson and DM emerge from a TeV-scale strongly-coupled sector as pNGBs, one key feature is  that the leading coupling between the SM and DM is provided by  higher-dimensional, derivative interactions with the Higgs field. The derivative Higgs portal mediates $s$-wave annihilation to SM particles, but leads to a  strong suppression of the DM scattering rate on ordinary matter. Thermal freeze-out can therefore yield the observed relic density for a DM mass of the order of $100 \, {\rm GeV}$, while the current severe limits of DM direct detection experiments are naturally evaded. Probes of composite pNGB~DM include indirect detection searches  and collider experiments. The collider reach on the derivative Higgs portal   has been recently analysed in vector-boson-fusion~(VBF) Higgs production~\cite{Ruhdorfer:2019utl}, finding a limited sensitivity at the LHC. This motivates studies of the indirect constraints on the derivative Higgs portal that arise from off-shell single-Higgs and on-shell double-Higgs production at hadron colliders~\cite{Haisch:2020ahr,inprep}. 

Besides the derivative Higgs portal, composite pNGB~DM models necessarily contain additional interactions   to provide a potential and Yukawa couplings for the Higgs boson and a mass for the DM candidate.  A theoretically motivated situation is one in which DM couples most strongly to the third generation of SM fermions. At the level of dimension-six operators, such interactions can either be of Yukawa type or involve  the product of a DM and a SM current. Detailed studies of the DM phenomenology of composite pNGB models where the Goldstone shift symmetry of DM is broken by the top or the bottom Yukawa coupling can be found in~\cite{Balkin:2017aep,Balkin:2018tma}. These analyses show that scenarios in which the shift symmetry is broken in the bottom sector are significantly less constrained by DM direct detection than those in which the top sector provides the leading symmetry breaking. In~composite pNGB models with sizeable DM-SM Yukawa couplings and a successful DM phenomenology, the leading $E_T^{\rm miss}$  signature is therefore expected to be DM production in association with bottom quarks. Unfortunately, this  process  can  only be  constrained poorly at the~LHC~\cite{Sirunyan:2017xgm,Aaboud:2017rzf,Aad:2021jmg}. If, on the other hand, effective current-current  interactions provide a relevant portal between the dark and the visible sector, large DM-top couplings are compatible with both  the bounds from DM (in)direct detection and the observed relic abundance if DM is sufficiently heavy~\cite{Ruhdorfer:2019utl}. As a result, such composite pNGB~DM models can  be tested at the LHC by searching for DM production in association with top-quark pairs $\big(t \bar t + \etmiss\big)$ or a top quark and a~$W$~boson~$\big(t W+ \etmiss\big)$. These mono-$X$ channels, from now on referred to as $t X + \etmiss$, have received a lot of attention from  the DM collider community~\cite{Lin:2013sca,Buckley:2014fba,Haisch:2015ioa,Arina:2016cqj,Haisch:2016gry,Sirunyan:2017xgm,Aaboud:2017rzf,Sirunyan:2017leh,Sirunyan:2018dub,Haisch:2018bby,Sirunyan:2019gfm,Aad:2020aob,Aad:2021hjy}.  

The main goal of this article is to analyse the LHC reach of the $t X+ \etmiss$ channels and to  constrain the parameter space of composite pNGB~DM models. To keep our discussion as model-independent as possible we will work in an effective field theory focusing on  the subset of  operators that lead to DM production in association with top quarks. Through loops such operators also lead to a $j + \etmiss$ signal, and we study  the limits on the parameter space of the pNGB~DM effective field theory  that are imposed by the corresponding mono-jet searches. We then  offer a comprehensive discussion of the phenomenological features of pNGB~DM models, including an analysis of the DM direct and indirect detection constraints as well as of the physics~of thermal freeze-out. The~search strategies and pNGB~DM benchmark models that we discuss are meant to set the stage  for dedicated experimental analyses by  ATLAS and CMS.  

Our work is organised as follows. In Section~\ref{sec:framework} we describe the structure of the  composite pNGB~DM models that we consider. Our Monte~Carlo~(MC) generation and our detector simulation are spelled out in~Section~\ref{sec:mcdetector}, while Section~\ref{sec:analysis} describes the analysis strategies to search for the relevant mono-$X$ signals. In~Section~\ref{sec:collider}~we  examine the sensitivity of the studied pNGB~DM signatures at upcoming LHC runs.  The present and future constraints on the pNGB~DM effective field theory  that arise from invisible Higgs decays are discussed in Section~\ref{sec:higgsinv}. The relevant non-collider limits are presented in Section~\ref{sec:noncollider}. We discuss our main results and give an outlook in Section~\ref{sec:conclusions}.  The impact of the assumed  systematic background uncertainties on our $t X+\etmiss$ projections is studied  in the supplementary material that can be found in Appendix~\ref{app:morenumerics}.

\section{Theoretical framework}
\label{sec:framework}

Throughout this article we will consider  theories in which both the SM Higgs doublet~$H$ and the DM candidate~$\chi$ arise as light pNGBs from a strongly-coupled sector. The DM candidate is a singlet under the SM gauge group and we assume it  to be a complex scalar. The terms of the  interaction Lagrangian relevant for the further discussion can be written as~\cite{Ruhdorfer:2019utl}
\beq \label{eq:LEFT}
\begin{split}
{\cal L}_{\chi H} & = \frac{c_d}{f^2} \hspace{0.25mm} \partial_\mu |\chi|^2 \hspace{0.25mm} \partial^\mu  |H|^2  -  \lambda \, |\chi|^2 |H|^2  \,, \\[2mm] 
{\cal L}_{\chi \psi} & = \frac{|\chi|^2}{f^2} \left ( c_t \hspace{0.25mm} y_t \hspace{0.25mm} \bar q_L \tilde{H} t_R + {\rm h.c.}\right ) + \frac{i}{f^2} \hspace{0.5mm} \chi^\ast   \overset{\leftrightarrow}{\partial_\mu} \hspace{0.5mm} \chi \sum_{\psi = q_L, t_R, b_R} d_\psi \hspace{0.25mm} \bar \psi \gamma^\mu \psi \,. 
\end{split}
\eeq
Here the terms in ${\cal L}_{\chi H}$ correspond to the derivative and marginal Higgs portal, respectively, while the terms in ${\cal L}_{\chi \psi}$ correspond to the  Yukawa-type DM-top coupling and the current-current type interactions between DM and the third-generation SM~quarks, respectively.  The common decay constant of the pNGBs is denoted by $f$, while the coefficients $c_i$, $\lambda$ and   $d_j$ are~$O (1)$  constants that we assume to be real such that CP is conserved. In~(\ref{eq:LEFT})~we have furthermore used the definition $\chi^\ast   \overset{\leftrightarrow}{\partial_\mu} \hspace{0.5mm} \chi = \chi^\ast \partial_\mu \chi -   \chi \partial_\mu \chi^\ast$,   and $q_L = (t_L, b_L)^T$ denotes the left-handed third-generation quark  doublet, $t_R$ ($b_R$) is the right-handed top-quark (bottom-quark) singlet,  $y_t = \sqrt{2} m_t/v$ is the top Yukawa coupling with $m_t \simeq 163 \, {\rm GeV}$ the top mass and $v \simeq 246 \, {\rm GeV}$ the Higgs~vacuum expectation value~(VEV), and we have defined $\tilde H^i = \varepsilon_{ij} \hspace{0.5mm} \big ( H^j \big)^\ast$ with~$\varepsilon_{ij}$ totally  antisymmetric and~$\varepsilon_{12}=1$. Notice that the current-current type operator in~${\cal L}_{\chi \psi}$ is absent if  hidden-charge conjugation (i.e.~$\chi \to -\chi^\ast$ and  $\psi \to \psi$) is preserved as in all explicit pNGB~DM models studied in~\cite{Balkin:2018tma}.  Moreover, this operator vanishes trivially if the DM candidate is a real scalar.  

Besides the four types of interactions introduced in~(\ref{eq:LEFT}), the full pNGB~DM effective field theory  can contain additional dimension-six operators such as  $ \chi^\ast   \overset{\leftrightarrow}{\partial_\mu} \hspace{0.5mm} \chi \hspace{0.25mm} \partial_\nu B^{\mu \nu}$ and  $|\chi|^2 \hspace{0.25mm} V_{\mu \nu} \hspace{0.25mm} V^{ \mu \nu}$. Here $V_{\mu \nu} = B_{\mu \nu}, W^i_{\mu \nu},   G^a_{\mu \nu}$ denotes the $U(1)_Y$, $S\!U(2)_L$ and $S\!U(3)_C$ field-strength tensor, respectively.  Since the latter two types of operators do not lead to  a relevant $t X + \etmiss$ signal at tree~level, such terms are not directly testable in DM production in association with top quarks. In~contrast, the presence of DM couplings with gauge bosons may have an important impact  on the calculation of the  DM (in)direct detection bounds and on the derivation of the DM relic density. To highlight the complementarity of collider and non-collider bounds in a simple fashion, we therefore restrict our analysis to the subclass of  models in which the leading effects  at the scale  at which DM and the Higgs boson emerge as composite pNGBs are well captured by the effective Lagrangians ${\cal L}_{\chi H}$ and ${\cal L}_{\chi \psi}$. However, we will  discuss and include pNGB~DM interactions with gauge bosons that are generated from~(\ref{eq:LEFT}) once radiative corrections are included, whenever these yield significant contributions (see Section~\ref{sec:noncollider}). 

We finally mention that under the assumption that the cancellation of gauge anomalies only depends on the SM fermion representations and not on the  structure of the pNGB~DM effective field theory  $\big($in particular the coefficients $d_\psi$ in~(\ref{eq:LEFT})$\big)$,  the current-current  type DM-top operator does not lead to~a~$j + \etmiss$~signal. In practice this requires one to introduce local  counterterms that cancel the anomalous contributions in the  five-point diagrams like the one shown on   the right-hand side in~Figure~\ref{fig:monoX2} --- see~\cite{Durieux:2018ggn,Bonnefoy:2020tyv,Feruglio:2020kfq} for related discussions of gauge anomalies in the context of the so-called SMEFT. Since we envisage that~(\ref{eq:LEFT}) describes new-physics scenarios in which the full SM gauge symmetry is preserved, a matching calculation in the full theory will always result in the required anomaly cancellation, and consequently a cancellation of the current-current  type  contributions to the mono-jet~signature for any value of the parameters $d_\psi$. 

\section{MC generation and detector simulation}
\label{sec:mcdetector}

\begin{figure}[!t]
\begin{center}
\includegraphics[width=0.65\textwidth]{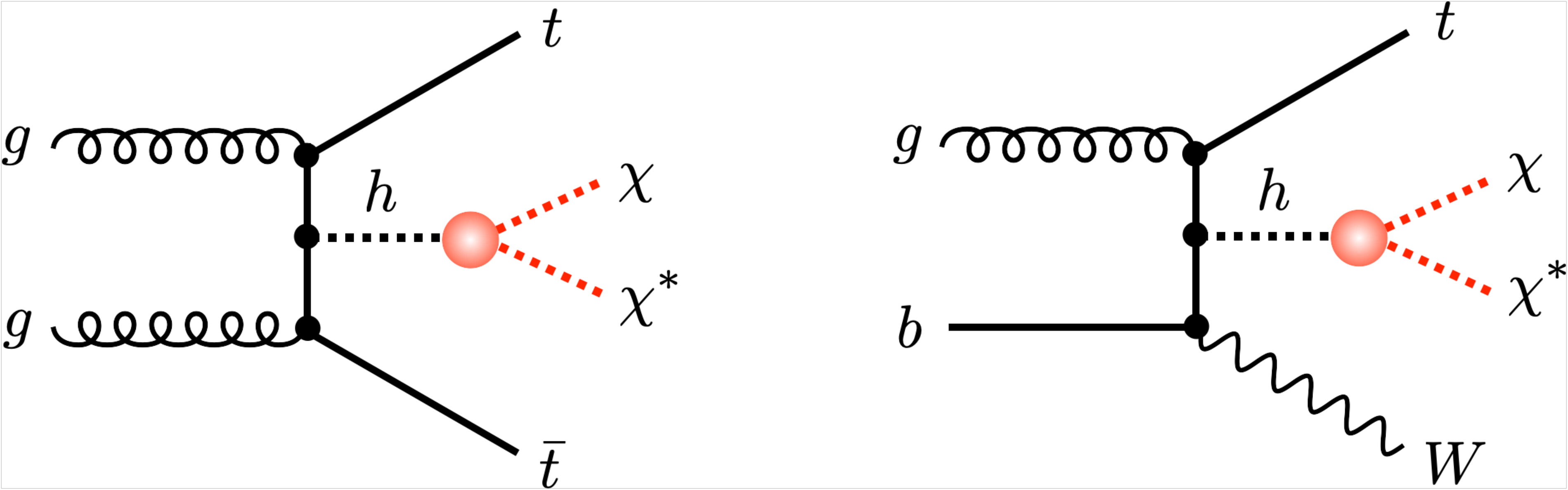}
\vspace{2mm} 
\caption{\label{fig:monoX1} Examples of diagrams with insertions of the DM-Higgs operators (filled red circles)  in~(\ref{eq:LEFT}) that lead to a $t \bar t +\etmiss$ (left) and $t W +\etmiss$ (right) signal.  The black dots indicate SM interactions.}
\end{center}
\end{figure}

\begin{figure}[!t]
\begin{center}
\includegraphics[width=0.985\textwidth]{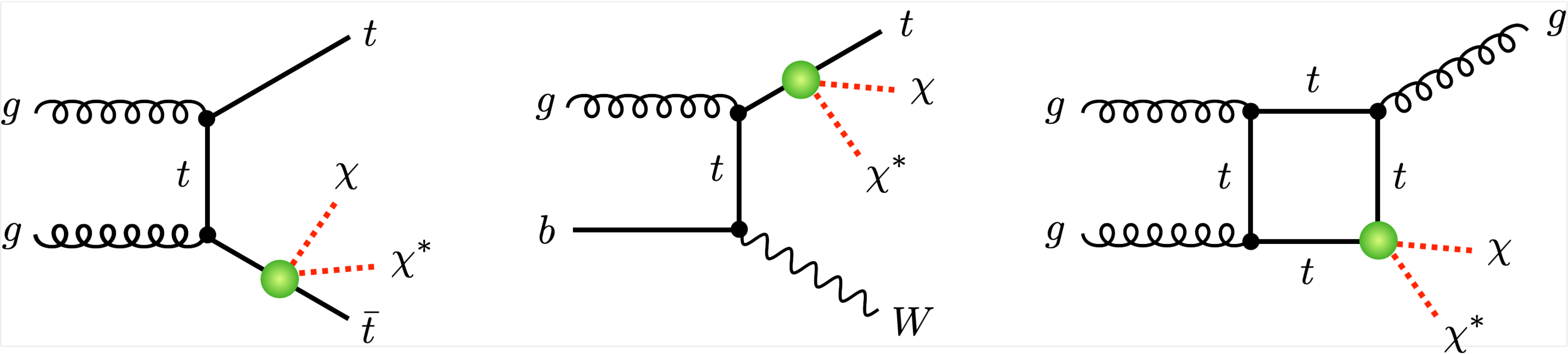}
\vspace{2mm} 
\caption{\label{fig:monoX2} Assortment of graphs with insertions of the DM-top operators (filled green circles) entering~(\ref{eq:LEFT}) that  give rise  to a $t \bar t +\etmiss$ (left), $t W +\etmiss$ (middle)  and $j + \etmiss$ (right) signature.}
\end{center}
\end{figure}

In our work we study the $t \bar t + \etmiss$, the $t  W + \etmiss$ and the  $j + \etmiss$ signatures that arise from insertions of the pNGB~DM operators introduced in~(\ref{eq:LEFT}). Examples of  leading-order~(LO) diagrams that involve DM-Higgs and DM-top operators are displayed in Figure~\ref{fig:monoX1} and~Figure~\ref{fig:monoX2}, respectively. Notice that only DM-top operators can lead to a LO mono-jet signal as illustrated by the graph shown on the right-hand side in Figure~\ref{fig:monoX2}. All our signal predictions assume proton-proton ($pp$) collisions at a  centre-of-mass~(CM) energy of $14 \, {\rm TeV}$ and are calculated using a {\tt FeynRules~2}~\cite{Alloul:2013bka}  implementation of the Lagrangian~(\ref{eq:LEFT}) in the {\tt UFO} format~\cite{Degrande:2011ua}. The~generation and showering of the mono-$X$ samples is  performed  with {\tt MadGraph5\_aMC@NLO}~\cite{Alwall:2014hca} at LO and {\tt PYTHIA~8.2}~\cite{Sjostrand:2014zea}, respectively, using  {\tt NNPDF3.0} parton distribution functions~(PDFs)~\cite{Ball:2014uwa}. In order to preserve both spin correlations and finite-width effects, final-state top quarks and $W$ bosons are decayed with~{\tt MadSpin}~\cite{Artoisenet:2012st}.

In the case of the  $t X+ \etmiss$ signatures, all SM processes that contain at least two charged leptons~($\ell = e, \mu$) coming from the decay of an  EW gauge boson $V = W,Z$ are included in the background simulation. We do not consider backgrounds  with either fake electrons from jet misidentification or with real non-isolated leptons from the decay of heavy-flavoured hadrons. A reliable estimate of these backgrounds depends on a detailed simulation of detector effects beyond the scope of this~article.  For the most recent ATLAS analyses involving leptonic final states~\cite{Aad:2021hjy, Aad:2020aob}, the background from non-prompt leptons is a few percent of the total background. The backgrounds from $\ttbar$~\cite{Campbell:2014kua}, $tW$~\cite{Re:2010bp}, $WW$, $WZ$ and $ZZ$ production~\cite{Melia:2011tj,Nason:2013ydw} are all generated at the next-to-leading order~(NLO) in~QCD with {\tt POWHEG~BOX}~\cite{Alioli:2010xd}. The $V + {\rm jets}$ backgrounds are generated at LO using {\tt  MadGraph5\_aMC@NLO} and include up to four additional jets. {\tt  MadGraph5\_aMC@NLO} is also used to simulate the $t\bar t V$ backgrounds with a multiplicity of up to two jets, while the $tZ$ and $tWZ$ backgrounds are obtained at LO with the same MC generator.  All partonic events are showered with~{\tt PYTHIA~8.2}. The samples produced with {\tt POWHEG~BOX} are normalised to the corresponding NLO QCD cross sections, except for~$t\bar{t}$, which is normalised to the  cross section obtained at the next-to-next-to-leading order~(NNLO) in~QCD plus next-to-next-to-leading logarithmic QCD corrections~\cite{Czakon:2011xx,Czakon:2013goa}. The~$V + {\rm jets}$ samples are normalised to the NNLO QCD cross sections~\cite{Anastasiou:2003ds,Gavin:2012sy} and the~$\ttbar V$~samples are normalised to the NLO QCD cross section as calculated by  {\tt  MadGraph5\_aMC@NLO}. 

For the $j + \etmiss$  signature, the dominant SM backgrounds arise from $V +{\rm jets}$ production. The~only relevant process not included in the  $t X+ \etmiss$ backgrounds described above is the $Z+\mathrm{jets}$ channel followed by the decay $Z \to \nu \bar \nu$. Like in  the earlier works~\cite{Haisch:2018hbm,Haisch:2020xjd} the corresponding background is  generated at LO with {\tt  MadGraph5\_aMC@NLO}, and can contain up to two additional jets. The~generation is performed in slices of the vector-boson transverse momentum~($p_T$), and the resulting events are showered with~{\tt PYTHIA~8.2} employing   a Catani-Krauss-Kuhn-Webber jet matching procedure~\cite{Catani:2001cc}. The inclusive signal region IM3  of the ATLAS~analysis~\cite{ATLAS:2021kxv} requires $\etmiss > 350 \, {\rm GeV}$, and  for these selections the background from $V +{\rm jets}$ production amounts to around 95\% of the total SM background.  The~$V +{\rm jets}$  samples are normalised such that the different contributions match the number of events in the IM3 signal region as estimated by ATLAS  scaled from  a  CM energy of $13 \, {\rm TeV}$ to~$14 \, {\rm TeV}$ and to the appropriate integrated luminosity. The additional minor backgrounds from~$t\bar{t}$, $tW$ and diboson production are the same as in the $t X + \etmiss$ case.

The actual physics analyses use experimentally identified electrons, muons, photons, jets~($j$) and~$\etmiss$. These objects are constructed from the stable particles in the generator output.  Jets~are built out of the momenta of all the stable particles  depositing energy in the calorimeter except for muons using the anti-$k_t$ algorithm~\cite{Cacciari:2008gp} with a radius parameter of $R=0.4$, as implemented in~{\tt FastJet}~\cite{Cacciari:2011ma}.  Jets originating from the hadronisation of bottom quarks ($b$-jets)  are experimentally identified~(i.e.~$b$-tagged) with high efficiency. The $\vec{p}_T^{\, {\rm miss}}$ vector with magnitude $\etmiss$  is constructed from  the transverse momenta of all the  invisible particles in the event. Detector effects are simulated by smearing the momenta of the analysis objects and by applying efficiency factors where applicable. The used smearing and efficiency functions  are tuned to reproduce the performance of the  ATLAS detector~\cite{Aad:2008zzm,Aad:2009wy}.  In particular,  the performance of the ATLAS $b$-tagging algorithm is taken from~\cite{Aad:2019aic}. For the mono-$X$ analyses  performed in this article,  a $b$-tagging working point is chosen that yields a  $b$-tagging efficiency of 77\%,  a  $c$-jet rejection  of~5~and a light-flavour jet rejection  of~110. More details on our detector simulation can be found in the earlier papers~\cite{Haisch:2016gry, Haisch:2018djm}.

\section{Mono-$\bm{X}$ analysis strategies}
\label{sec:analysis}

Below  we describe the analysis strategies to target the $t X+ \etmiss$ and $j + \etmiss$ signals that are due to the interactions described by~(\ref{eq:LEFT}).   For each analysis strategy we define the signal regions, spell out all  selection criteria and quantify the systematic uncertainties that plague the search strategy in~question. 

\subsection[$t X + \etmiss$  final states]{$\bm{t X + \etmiss}$ final states} 
\label{sec:tXetmissanalysis} 

The considered signal events include the decays of two $W$ bosons. We address the final states where only one or both of the $W$ bosons decay into charged leptons, which hereafter will be called semileptonic or fully-leptonic, respectively. Our $t X + \etmiss$ analysis is based on the definition of three orthogonal signal regions. The~first two signal regions target the associated production of a~$t \bar t$ pair and DM with SR1~(SR2) selecting semileptonic (fully-leptonic) events. The third signal region called  SR3   instead considers the associated production of a top quark,  a $W$ boson and DM, which is searched for in fully-leptonic events. The corresponding final states therefore involve  a single isolated charged lepton and two $b$-tagged jets~(SR1),  two isolated charged leptons  and two $b$-tagged  jets~(SR2)  or  two isolated charged leptons and a single $b$-tagged~jet~(SR3). Notice that $tW+\etmiss$ production typically has a smaller cross section than $t \bar t + \etmiss$ production.  However, in the case of the two-lepton final state, it has been shown in~\cite{Haisch:2018bby} that it is possible to devise a selection strategy that combines  the $t \bar t +\etmiss$ and the $t \bar W +\etmiss$ channels and has a significantly larger sensitivity than $t \bar t +\etmiss$ alone. Such a selection is based on the observation that events produced by a fully-leptonic $t \bar t$ decay contain two $\ell b$  pairs for both of which  the invariant mass~$m_{\ell b}$ is bounded
from above by $\sqrt{m_t^2 - M_W^2 } \simeq 153 \, {\rm GeV}$. This is not the case for the $tW$ production which contains only one $\ell b$ pair satisfying this bound. The two processes can thus be separated by defining the variable
\beq \label{eq:mblt}
m_{b \ell}^t =\mathrm{min} \hspace{0.5mm} \Big  (\mathrm{max} \hspace{0.5mm}  \big ( m_{\ell_1 j_a}, m_{\ell_2 j_b} \big  ) \Big ) \,,
\eeq
and putting a cut on $m_{b \ell}^t$  of around $160 \, {\rm GeV}$ to separate $t \bar t$ from $tW$ events. In~(\ref{eq:mblt}) the variables~$m_{\ell_1 j_a}$ and $m_{\ell_2 j_b}$ denotes the invariant mass of the leading and subleading leptons $\ell_1$ and $\ell_2$ and the  jets $j_a$ and $j_b$. The minimisation with respect to the jet pairs~$j_a$ and  $j_b$ runs over all of the $b$-tagged jets if the number  of $b$-tagged jets satisfies $N_b \geq 3$ or over the $b$-tagged jets and the untagged jet with the highest $b$-tagging weight if $N_b\leq2$.  Since the three signal regions are designed to have no events in common, the final search sensitivity of the $tX+\etmiss$ channel will be calculated after the statistical combination of SR1, SR2 and SR3. The~selection criteria corresponding to the three signal regions are summarised in Tables~\ref{tab:srdef1} and~\ref{tab:srdef23}.

\begin{table}
\def\arraystretch{1.25}
\begin{center}
\begin{tabular}{|l|c|}
\hline
Variable & SR1 selection    \\
\hline 
$N_{\ell}$ & $=1\,,$ \;  $p_{T}(\ell)>25 \, {\rm GeV}\,,$ \; $|\eta (\ell)|<2.5$ \\
$N_{j}$ & $\geq4 \,,$ \; $p_{T} (j)>(80,60,30,25) \, {\rm GeV}\,,$  \;$|\eta(j)| < 2.5$ \\
$N_{b}$ & $\geq2 \,,$ \; $p_{T} (b)>(80,25) \, {\rm GeV}\,,$ \; $|\eta(b)| < 2.5$ \\
\hline
$\etmiss$ & $> 550 \, {\rm GeV}$ \\
$m_T^\ell$ & $>180 \, {\rm GeV}$ \\
Topness & $>8$ \\
$m_{\rm top}^{\rm reclustered}$ &  $>150 \, {\rm GeV}$ \\
$H_{T, {\rm sig}}^{{\rm miss}}$ & $>15$ \\
$|\Delta\phi _{\ell, {\rm miss}} | $ & $>1.3$  \\
$|\Delta\phi _{\rm min}|$ & $>0.9$   \\
$|\Delta\phi_{bb}|$  & $<2.5$  \\
\hline
\end{tabular}
\vspace{4mm}
\caption{Definition of the signal region SR1. The number of charged leptons, light-flavoured jets and $b$-tagged jets are denoted by $N_\ell$, $N_j$ and $N_b$, respectively. For further details consult the text.}
\label{tab:srdef1}
\end{center}
\end{table}

In the case of SR1 the selection requirements are similar to the ones imposed in the signal region DM of~\cite{Aad:2020aob}. However, some variables have  been modified and  the values of the cuts have been optimised to our MC simulations of both the signal and the background at the high-luminosity upgrade of the LHC~(HL-LHC). The basic selection requires one and only one isolated charged lepton and at least four jets of which exactly two must be tagged as $b$-jets. Furthermore, jets tagged as hadronic decays of a $\tau$ lepton are vetoed. The employed cuts on the $p_T$ and pseudorapidities~($\eta)$ of the leptons and jets can be found in Table~\ref{tab:srdef1}. After the initial selections the dominant background is~$t \bar t$~production with one top quark decaying leptonically and the other one decaying hadronically. This background is strongly reduced by demanding $\etmiss > 550 \, {\rm GeV}$ and requiring a lower limit of $180 \, {\rm GeV}$ on the transverse mass of the charged lepton  defined as 
\beq  \label{eq:mTl}
m_T^\ell =  \sqrt{2 \hspace{0.5mm} |\vec{p}_T (\ell) | \hspace{0.5mm} | \vec{p}_T^{\, {\rm miss}}| \left (1-\cos \Delta\phi _{\ell, {\rm miss}} \right )} \,.
\eeq
Here $\vec{p}_T (\ell)$ denotes the components of the lepton momentum transverse to the beam, $\vec{p}_T^{\, {\rm miss}}$ is the vector sum of the transverse momenta of the invisible particles and $ \Delta\phi _{\ell, {\rm miss}}  = \Delta\phi ( \vec{p}_T (\ell)  , \vec{p}_T^{\, {\rm miss}} )$ is the azimuthal angular  separation between these two vectors. To reject events which are  incompatible with  top-quark decays, selections on the variables $\rm topness$~\cite{Graesser:2012qy} and $m_{\rm top}^{\rm reclustered}$~\cite{Aad:2020aob} are imposed. An additional rejection of the SM background  is achieved with selections on  $H_{T, {\rm sig}}^{\rm miss}$,~i.e.~the ratio of \etmiss built as the vector sum of the momenta of all the signal  jets and leptons in the event, reduced by $100 \, {\rm GeV}$ and divided  by its experimental resolution~\cite{ATLAS:2014mmo,ATLAS:2018uid}. Finally, cuts on the azimuthal angular separations $ \Delta\phi _{\ell, {\rm miss}}$,  $\Delta\phi_{\rm min}$ between $\vec{p}_{T}(j) $ and $\vec{p}_T^{\, {\rm miss}}$ for the four leading jets and  on  $\Delta\phi_{bb}$ between the two $b$-tagged  jets are imposed as detailed in Table~\ref{tab:srdef1}.

\begin{table}
\def\arraystretch{1.25}
\begin{center}
\begin{tabular}{|l|c|c|}
\hline
Variable &  SR2 selection &  SR3 selection  \\
\hline
$N_{\ell}$  & \multicolumn{2}{|c|}{$= 2\,,$ \; $p_{T}(\ell)>(25,20) \, {\rm GeV}\,,$ \; $|\eta (\ell)|<2.5$  }\\
$m_{\ell \ell}$  & \multicolumn{2}{|c|}{$>20 \, {\rm GeV}\,,$\;  $Z$-boson veto for OS leptons} \\
$N_{b}$ & \multicolumn{2}{|c|}{$\geq 1$, \;\; $p_{T} (b) >30 \, {\rm GeV}$, \;\; $|\eta ( b)| < 2.5$}\\
\hline
$m_{b \ell}^t$ &  $<160 \, {\rm GeV}$ & $>160 \, {\rm GeV}$ or $N_{j}=1$\\
$\etmiss$ & $>550 \, {\rm GeV}$ & $>350 \, {\rm GeV}$ \\
$|\Delta\phi_{min}|$ & n/a & $>0.8$ \\
$|\Delta\phi_{\rm boost}|$ & $<1.5$ & $<2.5$  \\
$M_{\rm scal}$ & n/a & $<500 \, {\rm GeV}$ \\
\mttwo  & $>100 \, {\rm GeV}$, shape fit  & $>170 \, {\rm GeV}$\\
\hline
\end{tabular}
\vspace{4mm}
\caption{As Table~\ref{tab:srdef1} but for the signal regions SR2 and SR3. More details can be found in the main text.}
\label{tab:srdef23}
\end{center}
\end{table}

The basis selection of events is common for the signal regions SR2 and SR3. It consists of the requirement of having exactly two isolated opposite-sign~(OS) leptons and the invariant mass of the  OS leptons has to fulfil $m_{\ell \ell} > 20 \, {\rm GeV}$. If the charged  leptons are of the same flavour, events with $71 \, {\rm GeV} < m_{\ell \ell} < 111 \,  {\rm GeV}$ are discarded  to suppress  backgrounds where the  lepton pair  arises from the decay $Z \to \ell^+ \ell^-$.  Furthermore, each event is required to contain at least one \btagged jet. The~relevant $p_T$ and $\eta$ selections of the OS leptons and $b$-jets are specified in Table~\ref{tab:srdef23}.  The~first selection that differs between the two signal regions is a cut on the $m_{b \ell}^t$ observable defined in~(\ref{eq:mblt}), which for SR2 (SR3) is  required to be smaller (larger) than $160 \, {\rm GeV}$. The variable $m_{b \ell}^t$ is only defined for events with at least two reconstructed jets and events with only one reconstructed jet  are assigned to SR3. Further selections are used to optimise the rejection of the SM backgrounds. In the case of SR2 (SR3) we require $\etmiss > 550 \, {\rm GeV}$ ($\etmiss > 350 \, {\rm GeV}$). The four leading jets furthermore have to satisfy $|\Delta\phi_{\rm min} |>0.8$ in the signal region SR3. The~variable~$\Delta\phi_{\rm boost}$ defined as the azimuthal angle difference between $\vec{p}_T^{\, {\rm miss}}$ and the vector sum of $ \vec{p}_T^{\, {\rm miss}}$, $\vec{p}_{T} (\ell_1)$ and $\vec{p}_{T} (\ell_2)$, must satisfy the requirement $|\Delta\phi_{\rm boost}|<1.5$ ($|\Delta\phi_{\rm boost}|<2.5$) for SR2 (SR3). In the case of the signal region SR3, we additionally demand that the scalar sum~$M_{\rm scal}$ of the transverse momenta of all the jets observed in the event satisfies $M_{\rm scal}<500\, {\rm GeV}$. Finally, in the signal region SR2 we require $\mttwo>100 \, {\rm GeV}$ and fit the shape of the \mttwo distribution (see for instance~\cite{Haisch:2018bby}), whereas for the signal region SR3 we impose the cut $\mttwo >170 \, {\rm GeV}$. Here~$\mttwo$ denotes the stransverse mass introduced in~\cite{Lester:1999tx}. 

Assuming an integrated luminosity of $3 \, {\rm ab}^{-1}$ at a CM energy of $14 \, {\rm TeV}$, the number of background events surviving the discussed requirements amounts  to 123, 34 and~48 in the case of SR1, SR2 and~SR3, respectively.  The signal efficiency depends on the DM mass and on the specific pNGB~DM model, and in the considered cases it is between a few tens of a percent and a few percent. Given the relatively large  number of surviving background events, the experimental reach  will  depend  sensitively on the systematic uncertainty  of the estimated SM backgrounds. The size of these uncertainties depends on the detector performance and the techniques used for the background evaluation, which are typically  based on a mixed MC and data-driven approach. Existing LHC analyses addressing signatures and a  phase space similar to our $tX+\etmiss$ strategy have  background uncertainties of  10\% to 30\%~$\big($see~\cite{Aaboud:2017rzf,Aad:2021hjy,Aad:2020aob}$\big)$.  In our numerical analysis we will assume a 15\% uncertainty on the backgrounds and  a 5\% uncertainty on the pNGB~DM signals. The latter uncertainty   should account for the effect of scale  variations and PDF uncertainties on the signal~modelling.

In addition to the analysis strategy described in detail above, we have also studied the sensitivity of the fully-leptonic signal regions SRt3 of~\cite{Aaboud:2017rzf} and ${\rm SR}^{\text{2-body}}$ of~\cite{Aad:2021hjy}, the semileptonic signal region DM of~\cite{Aad:2020aob} and the fully-hadronic signal regions SRt1 and SRt2 of~\cite{Aaboud:2017rzf} and SRA-TT of~\cite{Aad:2020sgw} to the parameter space of the pNGB~DM effective field theory . Our analyses rely in these cases on {\tt CheckMATE~2}~\cite{Dercks:2016npn}, which uses {\tt DELPHES~3}~\cite{deFavereau:2013fsa} as a fast detector simulation. We find that  for what concerns leptonic final states, the best limits on the parameters of~(\ref{eq:LEFT}) follow either from  the signal region DM or ${\rm SR}^{\text{2-body}}$, while  in the case of a fully-hadronic search the strategies SRt2 and SRA-TT fare equally well. It  furthermore turns out that the event selections employed in~\cite{Aaboud:2017rzf,Aad:2020aob,Aad:2020sgw,Aad:2021hjy}  perform at most as good but not  better than our  optimised $t X + \etmiss$  search strategy. We finally observe that for comparable sets of selection criteria the results from our parametrised simulation and the recast of the ATLAS analyses  are in good agreement which validates our simulation approach.

\subsection[$j+ \etmiss$ final state]{$\bm{j + \etmiss}$  final state}
\label{sec:jetmissanalysis}

In the case of the $j +\etmiss$  final state, the relevant pNGB~DM signal consists of a single high-transverse momentum jet and~$\etmiss$ associated to the production of a pair of DM particles. The signature  therefore resembles  the canonical mono-jet signal, which has received a significant amount of experimental~\cite{Aaboud:2016tnv,Aaboud:2017phn,Sirunyan:2017hci,ATLAS-CONF-2020-048} and theoretical~\cite{Lindert:2017olm} attention at the LHC, resulting in  high-precision estimates of the dominant $\etmiss$ backgrounds that are associated to  the production of an EW gauge boson accompanied by at least one high-transverse momentum jet. 

In our article we rely on the latest  ATLAS mono-jet analysis~\cite{ATLAS:2021kxv}. Specifically, we  employ $\etmiss>350 \, {\rm GeV}$ and require a high-transverse momentum jet with $p_T (j) >150 \, {\rm GeV}$ within $|\eta (j) |<2.4$, and no more than four jets with $p_T (j)>30 \, {\rm GeV}$ within $|\eta (j)|<2.8$. The selection~$|\Delta\phi_{\rm min}|>0.4$ is used to fully suppress the multi-jet background. All events containing a reconstructed electron or muon, or the hadronic decay  of a tau lepton are rejected. Our selection thus  closely resembles the signal region IM3 of~\cite{ATLAS:2021kxv}. The systematic uncertainty quoted by ATLAS in IM3  is 1.4\%, and we adopt this value as the systematic uncertainty on the total number of background events. Since we perform a multi-bin comparison of the shape of the $\etmiss$ variable, we also need to take into account uncertainties related to the~$\etmiss$  shape. For~each of the  $\etmiss$ bins considered in the analysis, ATLAS  gives an uncertainty which increases from around 1.4\% to 4\%  between $350 \, {\rm GeV}$ to $1.2 \, {\rm TeV}$. We apply these systematic uncertainties as bin-by-bin shape uncertainties in our $j + \etmiss$ analysis.  For the bins between $1.5 \, {\rm TeV}$ and $2 \, {\rm TeV}$ we furthermore assume an uncertainty of $5\%$, while we take an uncertainty of $8\%$ for the total number of events in the overflow bin with  $\etmiss > 2 \, {\rm TeV}$.  Notice that our uncertainty treatment corresponds to taking the uncertainties among different $\etmiss$ bins to be uncorrelated. In addition,  since the statistical uncertainties of the control regions, that are used to constrain the background, will get reduced with more luminosity, also the  systematic uncertainties are expected to decrease with larger data samples.  We thus believe that  our mono-jet study provides conservative results when applied to the full data set of the HL-LHC. 

\section{Constraints from $\bm{t X + \etmiss}$ and $\bm{j +\etmiss}$ searches at the LHC}
\label{sec:collider}

On the basis of the selection criteria given in Section~\ref{sec:analysis}, we will study the LHC sensitivity to the discussed mono-$X$ signatures. For each signature and each studied pNGB~DM benchmark, we evaluate the value of the cross section which  can be excluded at 95\%~confidence level~(CL) normalised to the nominal  LO cross section for the relevant model realisation as calculated by {\tt MadGraph5\_aMC@NLO}. The~experimental sensitivity is evaluated using a test statistic based on a profiled likelihood ratio and we make use of the CLs method~\cite{Read:2002hq} as implemented in  {\tt RooStats}~\cite{Moneta:2010pm}.

In Table~\ref{tab:tXMETsummary} we present the 95\%~CL bounds that derive from our $t X + \etmiss$ analysis for seven different DM masses in the range from $70 \, {\rm GeV}$ to $1 \, {\rm TeV}$. DM masses $m_\chi < m_h/2$ where $m_h \simeq 125 \, {\rm GeV}$ is the SM Higgs mass are not considered, because in this case  invisible Higgs decays generically represent the best way to probe pNGB~DM (see the discussion in Section~\ref{sec:higgsinv}). The shown limits correspond to the full data set of $3 \, {\rm ab}^{-1}$ that the HL-LHC is expected to collect at a CM energy of $14 \, {\rm TeV}$. Only   one free pNGB~DM effective field theory  parameter is allowed at a time. One observes  that HL-LHC~$t X +\etmiss$ searches are most sensitive to the  current-current type DM-fermion operators followed by the derivative Higgs portal operator and the Yukawa-type DM-top operator. The most difficult operator to probe  is the marginal Higgs portal, since it leads compared to the other pNGB~DM effective field theory  interactions in~(\ref{eq:LEFT}) to softer kinematic distributions, making a background suppression generically harder.  Notice that in the case of the marginal Higgs portal we have indicated the limits that correspond to a non-perturbative coupling, i.e. $|\lambda| > 4 \pi$,  by putting  parentheses around the corresponding results. We finally add that for $m_\chi = 1 \, {\rm TeV}$ the bounds on $f/\sqrt{|c_d|}$ and  $f/\sqrt{|c_t|}$  following from our $tX+\etmiss$ search strategy are so low that an effective field theory description might not be valid. The corresponding exclusion limits are therefore only indicative.

\begin{table} 
\def\arraystretch{1.25}
\begin{center}
\begin{tabular}{|c|ccccccc|}
\hline
\multicolumn{1}{|c|}{} & \multicolumn{7}{c|}{DM mass} \\ 
Parameter & $70 \, {\rm GeV}$ &  $100 \, {\rm GeV}$ &  $200 \, {\rm GeV}$ & $300 \, {\rm GeV}$ & $400 \, {\rm GeV}$ & $500 \, {\rm GeV}$ & $1 \, {\rm TeV}$ \\[1mm]
\hline 
$f/\sqrt{|c_d|}$ & $165 \, {\rm GeV}$ & $154 \, {\rm GeV}$ & $138 \, {\rm GeV}$ & $123 \, {\rm GeV}$ & $109 \, {\rm GeV}$ & $96 \, {\rm GeV}$ & $51 \, {\rm GeV}$ \\[1mm]
$|\lambda|$ & 2.4 & 6.0 & (23) & (55) &  (107) & (198) & (2315)  \\[1mm]
$f/\sqrt{|c_t|}$ & $153 \, {\rm GeV}$ & $150 \, {\rm GeV}$ & $137 \, {\rm GeV}$ & $122 \, {\rm GeV}$ & $107 \, {\rm GeV}$ & $96 \, {\rm GeV}$ & $50 \, {\rm GeV}$ \\[1mm]
$f/\sqrt{|d_{t_R}|}$ & $325 \, {\rm GeV}$ & $324 \, {\rm GeV}$ & $305 \, {\rm GeV}$ & $278 \, {\rm GeV}$ & $255 \, {\rm GeV}$ & $231 \, {\rm GeV}$ & $129 \, {\rm GeV}$ \\[1mm]
\hline
 \end{tabular}
\vspace{2mm}
\caption{95\%~CL bounds that derive from the $t X + \etmiss$ search strategy described in Section~\ref{sec:tXetmissanalysis} for seven different DM masses. All bounds assume $3 \, {\rm ab}^{-1}$ of integrated luminosity collected at a CM energy of $14 \, {\rm TeV}$. Only~the parameter shown in each line is taken into account, while all the remaining  couplings in~(\ref{eq:LEFT}) are set to zero. See text for further explanations.}
\label{tab:tXMETsummary}
\end{center}
\end{table}

\begin{table}
\def\arraystretch{1.25}
\begin{center}
\begin{tabular}{|c|ccccccc|}
\hline
\multicolumn{1}{|c|}{} & \multicolumn{7}{c|}{DM mass} \\ 
Parameter & $70 \, {\rm GeV}$ &  $100 \, {\rm GeV}$ &  $200 \, {\rm GeV}$ & $300 \, {\rm GeV}$ & $400 \, {\rm GeV}$ & $500 \, {\rm GeV}$ & $1 \, {\rm TeV}$ \\[1mm]
\hline 
$f/\sqrt{|c_t|}$ & $96 \, {\rm GeV}$ & $95 \, {\rm GeV}$ & $90 \, {\rm GeV}$ & $81 \, {\rm GeV}$ & $74 \, {\rm GeV}$ & $65 \, {\rm GeV}$ & $36 \, {\rm GeV}$ \\[2mm]
\hline
 \end{tabular}
\vspace{2mm}
\caption{As Table~\ref{tab:tXMETsummary} but for the $j + \etmiss$ search strategy described in Section~\ref{sec:jetmissanalysis}. }
\label{tab:jMETsummary}
\end{center}
\end{table}

The 95\%~CL bounds that  follow from our $j + \etmiss$ search strategy are collected in Table~\ref{tab:jMETsummary}. As~discussed at the end of Section~\ref{sec:framework}, mono-jet searches only allow to  test the Wilson coefficient~$c_t$ of the Yukawa-type DM-top operator in~(\ref{eq:LEFT}). It is evident from the shown results that the mono-jet bounds on $f/\sqrt{|c_t|}$ are not competitive with those obtained from $t X + \etmiss$.  We add that neglecting the uncertainty on the shape of the $\etmiss$ distribution~(see~Section~\ref{sec:jetmissanalysis}) in our $j + \etmiss$ analysis would improve the given  95\%~CL limits  by around 35\%. However, even then the mono-jet limits on $f/\sqrt{|c_t|}$  fall short of the bounds obtained from our $t X + \etmiss$ search strategy. Like in the case of the $tX+\etmiss$ bounds,  at high DM mass the $j+\etmiss$ limits should only be taken as indicative, because an effective field theory description may not be applicable in this regime. Benchmark scenarios with more than one non-zero pNGB~DM effective field theory  coefficient $c_i$, $\lambda$ and $d_j$ are  discussed in~Section~\ref{sec:conclusions}. 

\section{Constraints from invisible Higgs decays at the LHC}
\label{sec:higgsinv}

The terms in the first line of~(\ref{eq:LEFT}) will lead to  invisible Higgs decays at tree level if this process is kinematically allowed,~i.e.~for $m_\chi < m_h/2$. The relevant partial Higgs decay width reads 
\beq \label{eq:hDMDMwidth}
\Gamma \left ( h \to \chi^\ast \chi  \right ) = \frac{v^2}{16 \hspace{0.125mm} \pi \hspace{0.125mm} m_h} \, \sqrt{1 -  \frac{4 \hspace{0.25mm} m_\chi^2}{m_h^2} } \,  \left (\frac{m_h^2 \hspace{0.25mm} c_d }{f^2}  - \lambda \right )^2 \,,
\eeq
This formula can  be used to  translate experimental limits on the Higgs invisible branching ratio~${\rm BR} \left (h \to {\rm inv} \right)$ into constraints on $f/\sqrt{|c_d|}$ and $|\lambda|$. In fact, in the limit $m_\chi \ll m_h/2$ one obtains  the  95\%~CL exclusion limits 
\beq \label{eq:hinvpresent}
\frac{f}{\sqrt{|c_d|}} > 1.5 \, {\rm TeV} \,, \qquad |\lambda| < 7.2 \cdot 10^{-3}  \qquad (\text{LHC Run II}) \,, 
\eeq
by employing the best existing LHC bound of ${\rm BR} \left (h \to {\rm inv} \right) < 0.11$~\cite{ATLAS-CONF-2020-052}. At the HL-LHC it may be possible to set a limit on the Higgs invisible branching ratio of ${\rm BR} \left (h \to {\rm inv} \right) < 2.5 \cdot 10^{-2}$~\cite{Cepeda:2019klc}. This implies that the bounds~(\ref{eq:hinvpresent})  may be improved to 
\beq \label{eq:hinvfuture}
\frac{f}{\sqrt{|c_d|}} > 2.2 \, {\rm TeV} \,, \qquad |\lambda| < 3.3 \cdot 10^{-3}  \qquad (\text{HL-LHC}) \,.  
\eeq
Similar limits have also been given in~\cite{Ruhdorfer:2019utl}. Although the exclusion limits~(\ref{eq:hinvpresent}) and~(\ref{eq:hinvfuture}) have been derived under the assumption that either $c_d$~or~$\lambda$ is non-zero but not both, the obtained stringent limits indicate that invisible Higgs decays are the main  avenue to probe the pNGB~DM couplings~$c_d$~and~$\lambda$  for DM masses  $m_\chi < m_h/2$.  

\begin{figure}[!t]
\begin{center}
\includegraphics[width=0.5\textwidth]{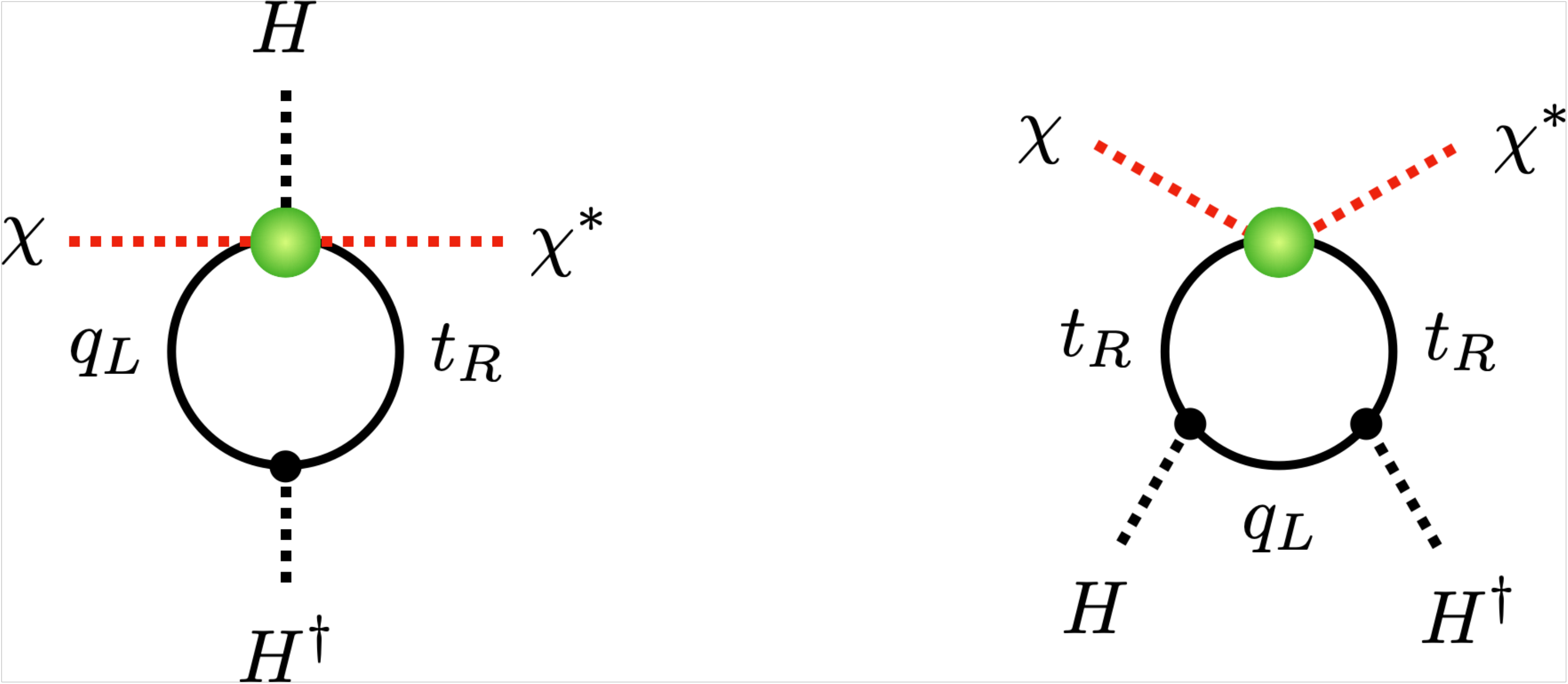}
\vspace{4mm} 
\caption{\label{fig:mixingdiagrams}   Left: An example of a diagram that describes the mixing of the Yukawa-type DM-top operator into the marginal Higgs portal operator. Right: Example graph that could lead to a mixing of the current-current  type DM-top operator into the DM-Higgs operators  in~(\ref{eq:LEFT}).  See text for further explanations.}
\end{center}
\end{figure}

At the loop level the  first interaction term in the second line of~(\ref{eq:LEFT}) can also lead to  invisible Higgs decays, because the  Yukawa-type DM-top operator mixes into the marginal Higgs portal operator through fermion loops --- see~the left Feynman diagram in Figure~\ref{fig:mixingdiagrams}. Assuming that the marginal Higgs portal coupling vanishes at the scale $\mu_f = O \left (f \right )$, we obtain the following  leading-logarithmic~(LL)  result 
\beq \label{eq:lambdaRGE}
\lambda = -\frac{3 \hspace{0.25mm} m_h^2 \hspace{0.25mm}  y_t^2 \hspace{0.25mm} c_t}{8 \hspace{0.25mm}  \pi^2 \hspace{0.25mm}  f^2} \, \ln \hspace{0.5mm} \frac{\mu_f}{\mu_h} \,,
\eeq
for the marginal Higgs portal coupling at the EW scale $\mu_h =  O \left (m_h \right)$. Notice that despite the fact that the contributions of the  Yukawa-type DM-top operator  to  the invisible  decays of the Higgs are loop suppressed the resulting constraints can still be important given the stringent bounds on ${\rm BR} \left (h \to {\rm inv} \right)$ that the HL-LHC is expected to  set. For instance, taking as an example  $c_t = 1$, $y_t \simeq 0.94$, $\mu_f = f$ and $\mu_h = m_h$, we find numerically that the  bound on $|\lambda|$ quoted in~(\ref{eq:hinvfuture}) leads to the limit 
\beq \label{eq:fyuk}
 f > 450 \, {\rm GeV}   \qquad(c_t = 1 \,, \text{HL-LHC}) \,, 
\eeq
on the  suppression scale of the Yukawa-type DM-top interactions introduced in~(\ref{eq:LEFT}).  In contrast to the Yukawa-type DM-top operator, the current-current type DM-quark operators do not mix into the DM-Higgs operators appearing in~(\ref{eq:LEFT}) since the sum over all one-loop diagrams of the type shown on the right-hand side of  Figure~\ref{fig:mixingdiagrams} vanishes. The pNGB~DM current-current type interactions therefore cannot be constrained by invisible Higgs decays even if $m_\chi < m_h/2$. 

\section{Constraints from DM  (in)direct detection and the relic density}
\label{sec:noncollider}

\begin{figure}[!t]
\begin{center}
\includegraphics[width=0.5\textwidth]{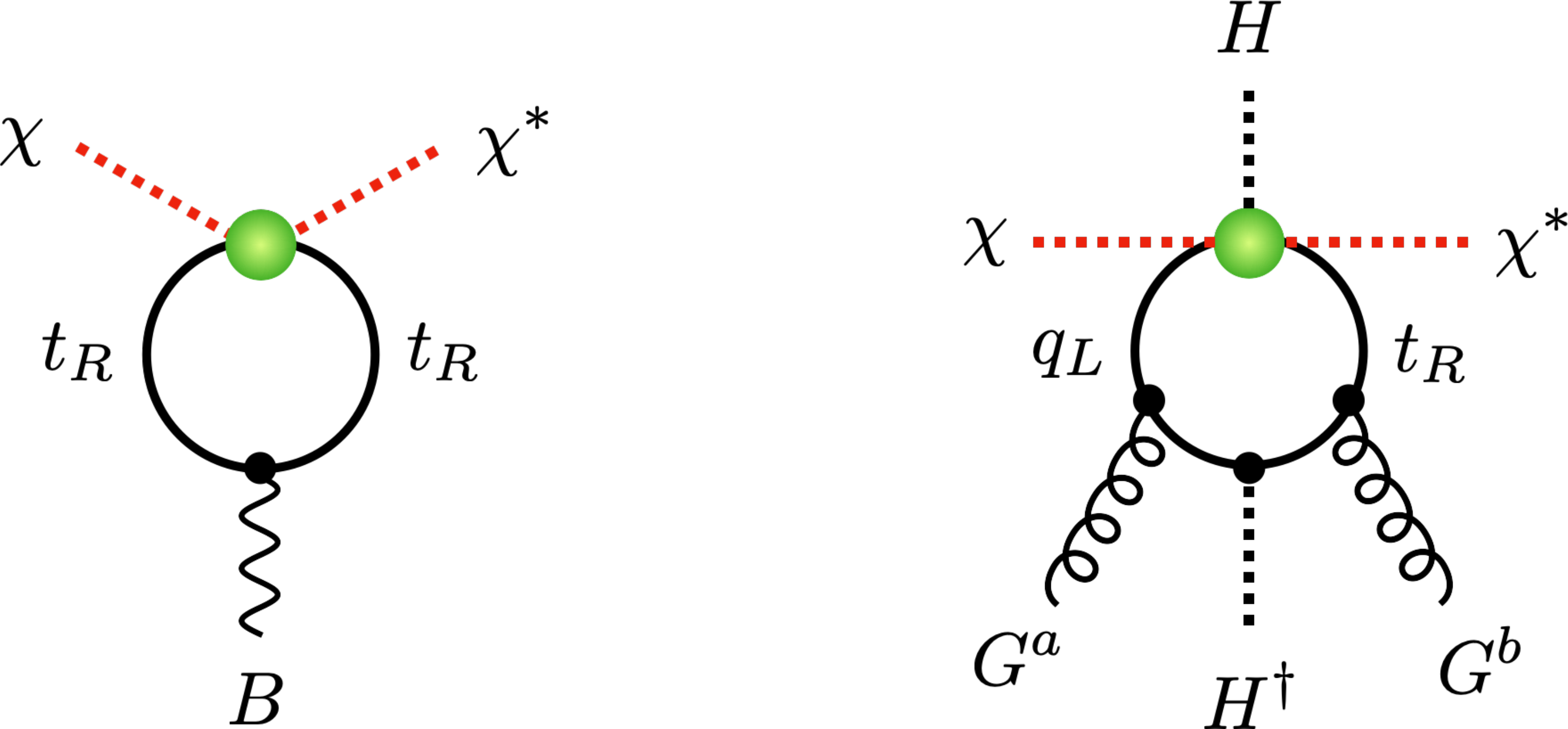}
\vspace{4mm} 
\caption{\label{fig:noncolliderdiagrams}   Left: Example diagram that describes the LL contribution of the current-current type DM-fermion operators  to  the Wilson coefficient of the DM-photon  operator appearing in~(\ref{eq:LDMV}). Right: A possible graph  involving the insertion of the Yukawa-type DM-top operator   that leads to a finite matching correction to the Wilson coefficient of the DM-gluon operator in~(\ref{eq:LDMV}). See text for further details.}
\end{center}
\end{figure}

Even under the assumption that the interactions  in~(\ref{eq:LEFT}) provide the leading new-physics effects at the scale $\mu_f$ at which the spin-$0$ fields emerge as composite pNGBs, the inclusion of radiative corrections can spoil this picture at the low energies probed  in DM-nucleon scattering or DM annihilation~(see~\cite{Hisano:2010ct,Freytsis:2010ne,Hisano:2011cs,Hill:2011be,Frandsen:2012db,Haisch:2013uaa,Hill:2013hoa,Crivellin:2014qxa,Crivellin:2014gpa,DEramo:2014nmf,DEramo:2016gos,Bishara:2018vix} for further examples of relevant loop corrections in DM interactions). In fact, in the case at hand, we find that loop diagrams like those displayed in Figure~\ref{fig:noncolliderdiagrams} induce couplings between DM and the $U(1)_Y$ gauge boson  or a pair of gluons. After EW symmetry breaking the DM gauge-boson interactions relevant  for DM-nucleon scattering can be cast into the~form 
\beq \label{eq:LDMV}
{\cal L}_{\chi V} = \frac{i \hspace{0.25mm} e \hspace{0.25mm} c_A}{16 \hspace{0.25mm}\pi^2  \hspace{0.25mm} f^2}  \hspace{0.5mm} \chi^\ast   \overset{\leftrightarrow}{\partial_\mu} \hspace{0.5mm} \chi \hspace{0.25mm} \partial_\nu F^{\mu \nu} +    \frac{ g_s^2  \hspace{0.25mm} d_G}{16 \hspace{0.25mm}\pi^2  \hspace{0.25mm} f^2} \hspace{0.5mm}   |\chi|^2 \hspace{0.25mm}  G_{\mu \nu}^a  G^{a, \mu \nu} \,, 
\eeq
where $e \simeq  0.3$ is the elementary electromagnetic charge, $g_s \simeq 1.2$ denotes the strong coupling constant and $F_{\mu \nu}$ represents the electromagnetic field strength tensor. The leading contributions to the Wilson coefficients of the operators in~(\ref{eq:LDMV})  read
\beq \label{eq:cBcG}
c_A = \frac{4}{3} \left ( d_{q_L} + 2 d_{t_R} - d_{b_R} \right )  \, \ln \hspace{0.5mm} \frac{\mu_f}{\mu_h} \,, \qquad 
d_G = -\frac{c_t}{3} \,. 
\eeq
Notice that the Wilson coefficient $c_A$ contains only the LL correction associated to operator mixing,  while the result for  $d_G$ corresponds to a finite matching correction obtained in the limit of  infinite top-quark mass. 

Including the tree-level contributions that arise from the marginal Higgs portal operator appearing in~(\ref{eq:LEFT}) as well as loop-induced interactions described by~(\ref{eq:cBcG}), the spin-independent~(SI) DM-nucleon cross section can be written as 
\beq \label{eq;sigmaSI}
\sigma_{\rm SI} = \frac{1}{\pi}  \left ( \frac{m_\chi  \hspace{0.25mm} m_N}{m_\chi + m_N} \right )^2 \frac{1}{A^2}  \hspace{0.25mm} \left \{  \hspace{0.5mm} \frac{A  \hspace{0.25mm}m_N}{2  \hspace{0.25mm}m_\chi} \left [ \left ( 1 - \frac{7  \hspace{0.25mm} f^{N}_{T_G}}{9} \right ) \frac{\lambda}{m_h^2}  - \frac{2  \hspace{0.25mm}f^{N}_{T_G} \hspace{0.125mm} d_G}{9  \hspace{0.25mm}  f^2}   \right ] + \frac{Z \hspace{0.25mm} e^2 \hspace{0.25mm} c_A}{16 \hspace{0.25mm} \pi^2\hspace{0.25mm}  f^2} \hspace{0.5mm}  \right  \}^2  \,.
\eeq
Here $A$ ($Z$) is the mass (atomic) number of the nucleus, $m_N \simeq 0.939 \, {\rm GeV}$ denotes the average nucleon mass and $f^{N}_{T_G} = 1 - \sum_{q=u,d,s} f_{T_q}^N \simeq 0.89$ is the effective gluon-nucleon coupling, and its numerical value corresponds to the values $f_{T_u}^N \simeq 0.019$, $f_{T_d}^N \simeq 0.045$ and $f_{T_s}^N \simeq 0.043$~\cite{Junnarkar:2013ac,Hoferichter:2015dsa} for the quark-nucleon matrix elements. Furthermore, notice that the contribution in~(\ref{eq;sigmaSI}) proportional to~$c_A$ arises from $t$-channel photon exchange and that the  corresponding  form factors simply count the number of valence quarks of the nucleons, i.e.~$f^p_{V_u} = f^n_{V_d} = 2$ and $f^p_{V_d} = f^n_{V_u} = 1$. 

For $m_\chi =100 \, {\rm GeV}$ the latest XENON1T  90\%~CL upper limit on the SI DM-nucleon cross section reads $\sigma_{\rm SI} < 9.12 \cdot 10^{-47} \, {\rm cm^2}$~\cite{Aprile:2018dbl}.  Using~(\ref{eq;sigmaSI}) with $A = 131$ and $Z = 54 $ for xenon, this bound can be readily translated into limits on the Wilson coefficients  of the relevant pNGB~DM operators  in~(\ref{eq:LEFT}). In the case of the marginal Higgs portal, we find  in agreement with~\cite{Ruhdorfer:2019utl} the 90\%~CL exclusion limit
\beq \label{eq:DDmar}
|\lambda| < 1.0 \cdot 10^{-2} \,.
\eeq 
Setting~$c_t = 1$ in~(\ref{eq:lambdaRGE}) and~(\ref{eq:cBcG})  as well as using  $\mu_f = f$ and  $\mu_h = m_h$, and setting $d_{q_L} = d_{t_R} = d_{b_R} =~1$ in~(\ref{eq:cBcG}) ,  we obtain in addition  the  lower bounds 
\beq \label{eq:DDyukcur}
\begin{split}
& f > 510 \, {\rm GeV}  \qquad \hspace{0.0mm} (c_t = 1) \,, \\[2mm]
& f > 1.3 \, {\rm TeV}  \qquad \hspace{1.5mm} (d_{q_L} = d_{t_R} = d_{b_R} = 1) \,, 
\end{split}
\eeq
on the suppression scale of the Yukawa-type and the current-current type DM-fermion interactions entering~(\ref{eq:LEFT}), respectively. Although we have  considered  in all  cases only the effect of one type~of pNGB~DM operator at the scale $\mu_f$ at a time, the limits~(\ref{eq:DDmar}) and~(\ref{eq:DDyukcur}) show that the null results of the DM direct detection experiments generically allow to set stringent bounds on the Wilson coefficients of the marginal Higgs portal and the pNGB~DM-fermion operators in~(\ref{eq:LEFT}). In contrast the derivative Higgs portal operator remains unconstrained by DM direct detection even after one-loop corrections are included in the calculation of the SI DM-nucleon cross section.

In order to understand the physics of DM indirect detection and thermal-freeze out in composite pNGB~DM models,  we first write the velocity-averaged cross section for annihilation of DM into a SM final state $X$ as 
\beq \label{eq:annxsec}
\left \langle  \sigma \left (\chi^\ast \chi \to X \right ) v \right  \rangle \left ( T \right ) =  a_X + T \hspace{0.25mm} b_X \,.
\eeq
Here $T$ denotes the DM temperature and thus the coefficient  $a_X$ ($b_X$) describes the $s$-wave ($p$-wave) contribution. Notice that in today's Universe $T_0 \simeq 0$, while at freeze-out $T_f \simeq m_\chi/25$. This means that the $p$-wave coefficient $b_X$ can usually be neglected in the calculation of the DM indirect detection constraints, while it can be relevant in the computation of the relic abundance $\Omega_\chi h^2$, in particular if the corresponding $s$-wave coefficient $a_X$ is parametrically suppressed. 

An example where such a parametric suppression is at work in the context of~(\ref{eq:LEFT}) is the annihilation of DM into a bottom-antibottom quark pair, i.e.~$\chi^\ast \chi \to b \bar b$. In this case, we find  that the relevant $s$-wave and $p$-wave coefficients are well approximated by 
\beq \label{eq:abbbbb}
a_{b \bar b}  \simeq \frac{3 \hspace{0.25mm} m_b^2}{4 \pi} \left | \hspace{0.25mm}  \frac{1}{4 \hspace{0.25mm} m_\chi^2 - m_h^2 + i \hspace{0.25mm} m_h \hspace{0.25mm} \Gamma_h} \, \left ( \frac{4  \hspace{0.25mm} m_\chi^2  \hspace{0.25mm}  c_d}{f^2}  - \lambda \right )  \hspace{0.25mm}  \right |^2  \,, \qquad 
b_{b \bar b}  \simeq \frac{3 \hspace{0.25mm} m_\chi}{8  \hspace{0.125mm} \pi}  \frac{d_{q_L}^{\hspace{0.125mm} 2} + d_{b_R}^{\hspace{0.125mm} 2} }{f^4} \,, 
\eeq
if  the DM mass is sufficiently above the bottom-quark threshold at $m_\chi = m_b \simeq 4.2 \, {\rm GeV}$. In~the above expression for $a_{b \bar b}$,  the total  decay width  of the Higgs boson including contributions from $h \to \chi^\ast \chi$~$\big($see~Section~\ref{sec:higgsinv}$\big)$ is denoted by $\Gamma_h$.    For $m_b < m_\chi \lesssim m_W$ with the $W$-boson mass $m_W\simeq 80.4 \, {\rm GeV}$, the $\chi^\ast \chi \to b \bar b$ channel generically provides  the dominant mechanism to set~$\Omega_\chi h^2$ in composite pNGB~DM models described by~(\ref{eq:LEFT}). In fact, it turns  out  that for $m_\chi \ll m_h/2$ the velocity  suppression of the $p$-wave contribution in~(\ref{eq:abbbbb}) is less severe than the bottom-mass~suppression of the $s$-wave contribution  in~(\ref{eq:abbbbb}). The current-current type DM-fermion operators introduced in~(\ref{eq:LEFT}) can therefore play an important role in thermal freeze-out for $m_\chi < m_h/2$.

For $m_\chi \gtrsim m_W$ the $\chi^\ast \chi \to W^+ W^-, ZZ, hh, t \bar t$ channels  dominate DM annihilation. These~processes all receive unsuppressed $s$-wave contributions, rendering the associated  $p$-wave contributions phenomenologically irrelevant. For DM masses sufficiently far above the EW~scale, we find the following approximations for the $s$-wave coefficients
\beq \label{eq:aVVahhatt}
\begin{split}
a_{X}  \simeq \frac{N_X \hspace{0.25mm} m_\chi^2}{4 \pi} \left [ \frac{c_d}{f^2} - \frac{\lambda}{4 \hspace{0.25mm} m_\chi^2} \right ]^2 \,, \qquad 
a_{t \bar t}  \simeq \frac{3 \hspace{0.25mm} m_t^2}{4 \pi} \left [ \frac{c_d+c_t}{f^2} - \frac{\lambda}{4 \hspace{0.25mm} m_\chi^2} \right ]^2  \,, 
\end{split}
\eeq
where $X = W^+ W^-, ZZ, hh$ and $N_{W^+W^-} = 2$, $N_{ZZ} = N_{hh} = 1$. The above results can be shown to agree with the calculations performed in~\cite{McDonald:1993ex} after taking the limit of large DM mass. Notice that in this limit, DM annihilation to $W$ and $Z$ bosons reduces to three times the contribution from annihilation to the Higgs boson, as expected in the $S\!U(2)_L \times U(1)_Y$ symmetric limit. Given that the size of the marginal Higgs portal coupling $\lambda$ is strongly constrained by DM direct detection~$\big($see~(\ref{eq:DDmar})$\big)$, the expressions~(\ref{eq:aVVahhatt}) also imply that in viable composite pNGB~DM models the derivative Higgs portal operator generically provides the dominant contribution to DM annihilation for $m_\chi \gg m_t$. As~a~result thermal freeze-out becomes a model-independent prediction in this limit, in the sense that the value of~$\Omega_\chi h^2$ to first approximation only depends on $m_\chi$ and $f/\sqrt{|c_d|}$.

\begin{figure}[!t]
\begin{center}
\includegraphics[width=0.9\textwidth]{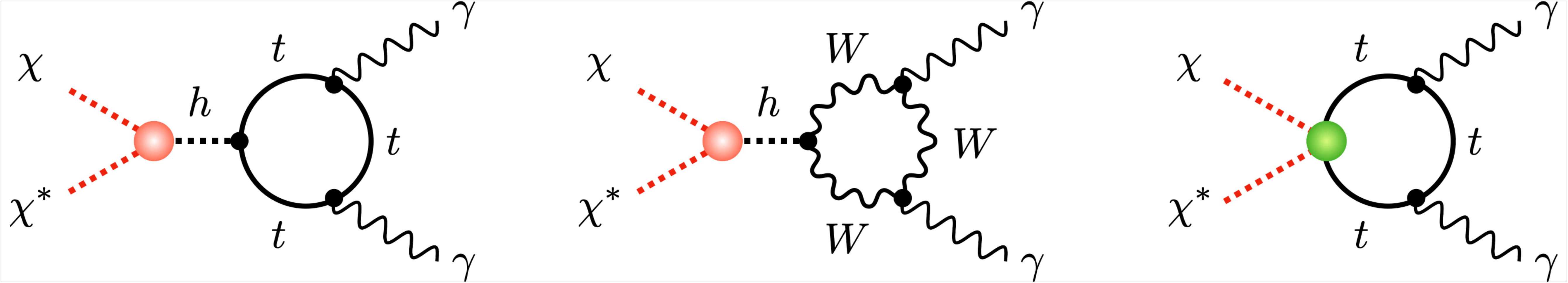}
\vspace{4mm} 
\caption{\label{fig:gammarays}    Example diagrams that lead to  the process $\chi^\ast \chi \to \gamma \gamma$ .  Further details can be found in the  text.}
\end{center}
\end{figure}

In addition to the DM annihilation channels discussed so far, DM annihilation into mono-chromatic photons can  provide a relevant indirect-detection signature in composite pNGB~DM models. As shown in Figure~\ref{fig:gammarays}, this signature receives two types of contributions. The first is associated to $s$-channel exchange of a Higgs boson with subsequent  decay of the Higgs into a pair of photons,~i.e.~$\chi^\ast \chi \to h \to \gamma \gamma$, and proceeds through the insertion of a DM-Higgs operator and a loop of top quarks (left diagram) or $W$~bosons (middle diagram). The corresponding form factors describing fermion and gauge-boson loops are given by  
\beq \label{eq:FpsiFV}
\begin{split}
F_{\psi} \hspace{0.25mm} (\tau) & = \frac{3 \hspace{0.25mm} \tau}{2} \left [  1 + \left (1 - \tau   \right ) \arctan^2 \frac{1}{\sqrt{\tau -1}}  \right ] \,, \\[2mm]
F_{V} \hspace{0.25mm} (\tau) & = \frac{1}{7} \left [ 2 +  3 \hspace{0.25mm} \tau + 3  \hspace{0.25mm}  \tau  \left (2 -  \hspace{0.25mm} \tau  \right )\arctan^2 \frac{1}{\sqrt{\tau -1}}  \right ] \,, 
\end{split}
\eeq
respectively, and are normalised such that $F_{\psi} \hspace{0.25mm}  (\infty) = F_{V} \hspace{0.25mm}  (\infty) = 1$.  The second type of contributions involves the insertion of the Yukawa-type DM-top operator introduced in~(\ref{eq:LEFT}) and leads directly to~the $\chi^\ast \chi  \to \gamma \gamma$ transition via a top-quark loop (right diagram in Figure~\ref{fig:gammarays}). Including both types of contributions, the $s$-wave coefficient corresponding to $\chi^\ast \chi \to \gamma \gamma$ annihilation can be written as 
\beq \label{eq:aAA}
a_{\gamma \gamma}  = \frac{\alpha^2  m_\chi^2}{8 \hspace{0.25mm}  \pi^3} \left |  \hspace{0.25mm} \frac{1}{4 \hspace{0.25mm} m_\chi^2 - m_h^2 + i \hspace{0.25mm} m_h \hspace{0.25mm} \Gamma_h} \left ( \frac{4  \hspace{0.25mm} m_\chi^2  \hspace{0.25mm}  c_d}{f^2}  - \lambda \right )  \left [ \frac{8 \hspace{0.25mm} F_\psi \hspace{0.25mm} (\tau_t)}{9}  - \frac{7 \hspace{0.25mm} F_V \hspace{0.25mm} (\tau_W)}{2}  \right ] +\frac{8 \hspace{0.25mm}  c_t}{9 f^2}  F_\psi \hspace{0.25mm} (\tau_t)  \hspace{0.25mm} \right |^2  \,, 
\eeq
where $\tau_i = m_i^2/m_\chi^2-i \varepsilon$ with $\varepsilon$  being a positive infinitesimal real number. Notice that the $s$-channel Higgs exchange contribution in~(\ref{eq:aAA}) is resonantly enhanced at  $m_\chi = m_h/2$, and as a result the DM indirect detection constraints from the observation of $\gamma$-ray lines are generically most stringent in the vicinity of the  Higgs pole. 

Based on~(\ref{eq:annxsec}) to~(\ref{eq:aVVahhatt}),  the present abundance of DM in the Universe is approximately given by the following formula 
\beq \label{eq:omegapocket}
\frac{\Omega_\chi h^2}{0.12} \simeq  \frac{3 \cdot 10^{-26} \, {\rm cm}^3/{\rm s}}{ \langle \sigma v \rangle_f} \,, \qquad 
\langle \sigma v \rangle_f = \frac{1}{2} \sum_{X} \left \langle  \sigma \left (\chi^\ast \chi \to X \right ) v \right  \rangle \big ( T_f \big ) \,,
\eeq
where the sum over $X$ involves all annihilation channels that are kinematically accessible at a given DM mass. Notice that the factor of $1/2$ in the definition of $\langle \sigma v \rangle_f$ takes into account that DM is not self-conjugate in our case. The same factor of $1/2$ appears when one calculates the $\gamma$-ray flux from the annihilation cross section~(\ref{eq:aAA}). While~(\ref{eq:omegapocket}) represents a useful expression to estimate~$\Omega_\chi h^2$, we will use~{\tt micrOMEGAs}~\cite{Belanger:2018ccd} in our numerical analysis of the constraints on the pNGB~DM parameter space following from the requirement to reproduce the  relic abundance of $\Omega_\chi h^2 = 0.120 \pm 0.001$ as measured by {\rm PLANCK}~\cite{Aghanim:2018eyx}. {\tt micrOMEGAs} is also used to determine the DM indirect detection exclusion limits. 

\section{Discussion}
\label{sec:conclusions}

\begin{figure}[!t]
\begin{center}
\includegraphics[width=0.8\textwidth]{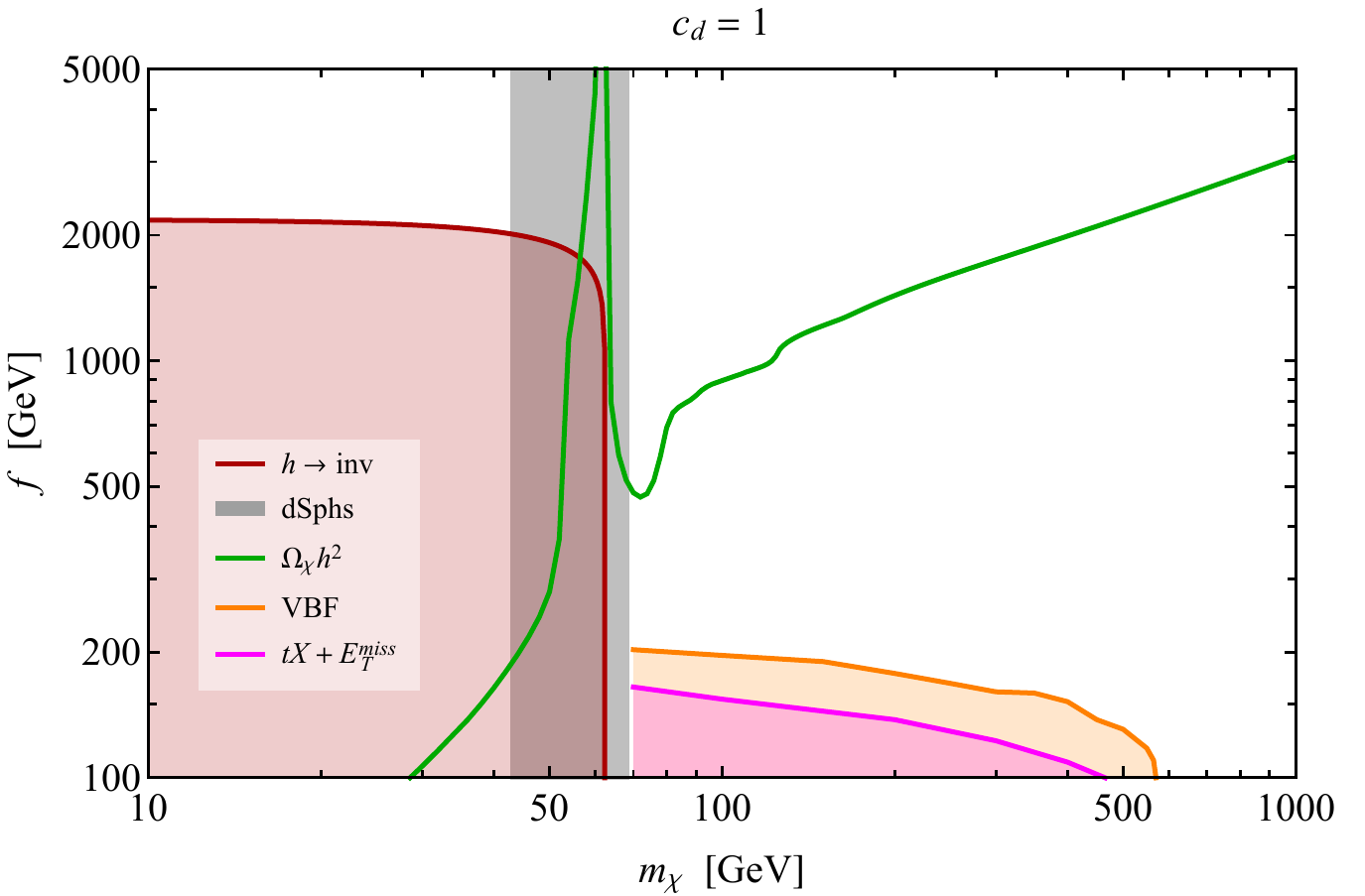} 
\vspace{2mm} 
\caption{\label{fig:summaryS1} Constraints in the $m_\chi \hspace{0.25mm}$--$\hspace{0.25mm} f$~plane for the derivative Higgs portal model. The pNGB~DM effective field theory  parameters not shown in the headline of the plot  are set to zero to obtain the displayed results. The~dark red region is excluded by the projected HL-LHC 95\%~CL limit on the Higgs invisible branching ratio of ${\rm BR} \left (h \to {\rm inv} \right) < 2.5 \cdot 10^{-2}$~\cite{Cepeda:2019klc}. The vertical grey band displays  the DM  mass range that is excluded at 95\%~CL by  the dSphs analysis of {\rm Fermi-LAT} and {\rm DES}~\cite{Fermi-LAT:2016uux} assuming $\chi^\ast \chi \to b \bar b$ annihilation. The green curve corresponds to the value $\Omega_\chi h^2 = 0.12$  of the DM relic density as determined by  {\rm PLANCK}~\cite{Aghanim:2018eyx}. In~the parameter space above the green curves the Universe is overclosed. The orange region indicates the 95\%~CL exclusion limit derived in~\cite{Ruhdorfer:2019utl} from a study of off-shell invisible Higgs production in the~VBF channel at the HL-LHC, while the magenta region represents the corresponding exclusion limit obtained by our  $t X + \etmiss$ search strategy. Consult the main text for further details. }
\end{center}
\end{figure}

\begin{figure}[!t]
\begin{center}
\includegraphics[width=0.8\textwidth]{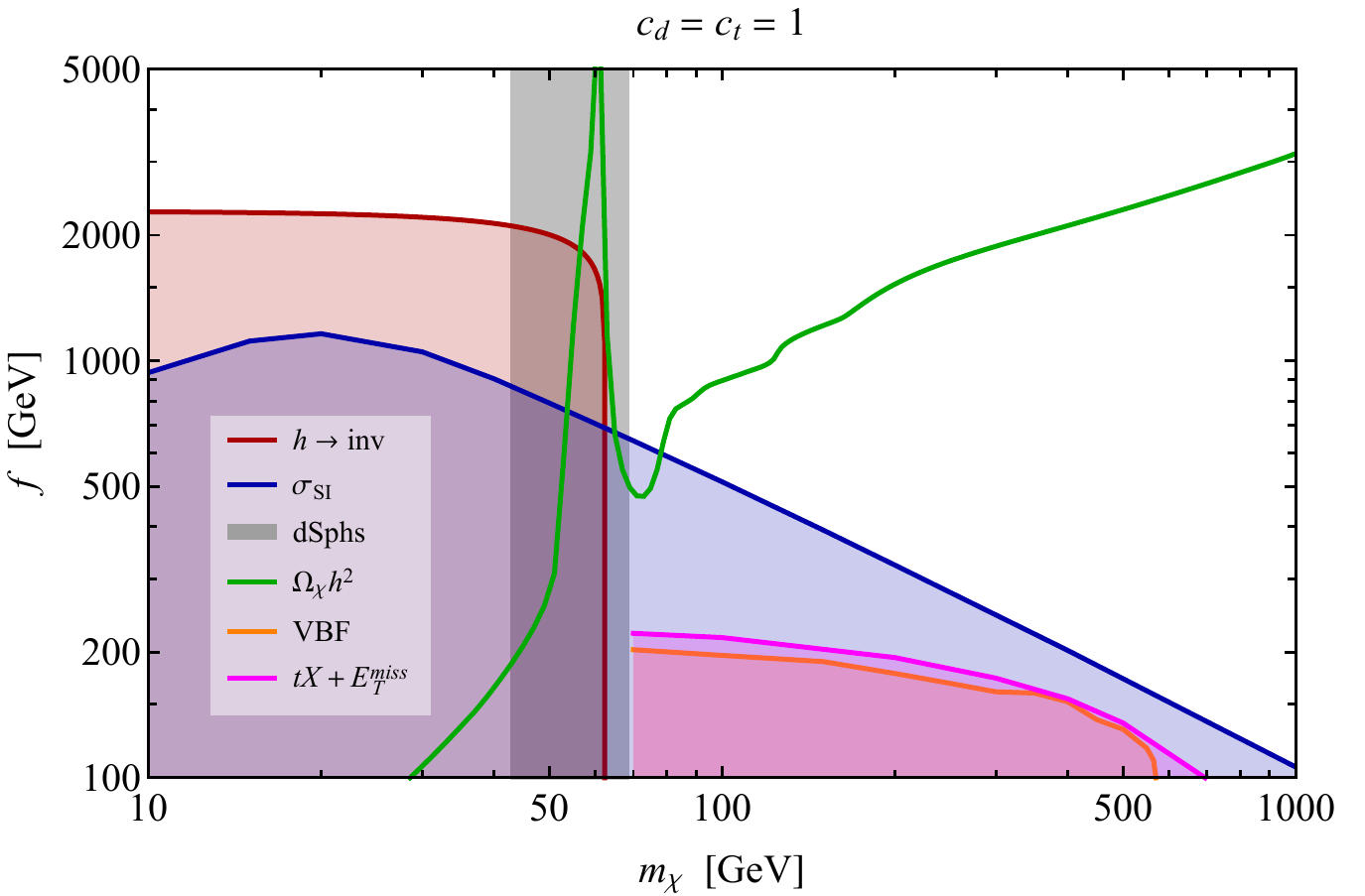} 
\vspace{2mm} 
\caption{\label{fig:summaryS2} As Figure~\ref{fig:summaryS1} but for the  pNGB~DM benchmark model with $c_d=c_t=1$. The blue region is excluded by  the 90\%~CL bound on the SI DM-nucleon cross section $\sigma_{\rm SI}$ as determined by XENON1T~\cite{Aprile:2018dbl}.}
\end{center}
\end{figure}

\begin{figure}[!t]
\begin{center}
\includegraphics[width=0.8\textwidth]{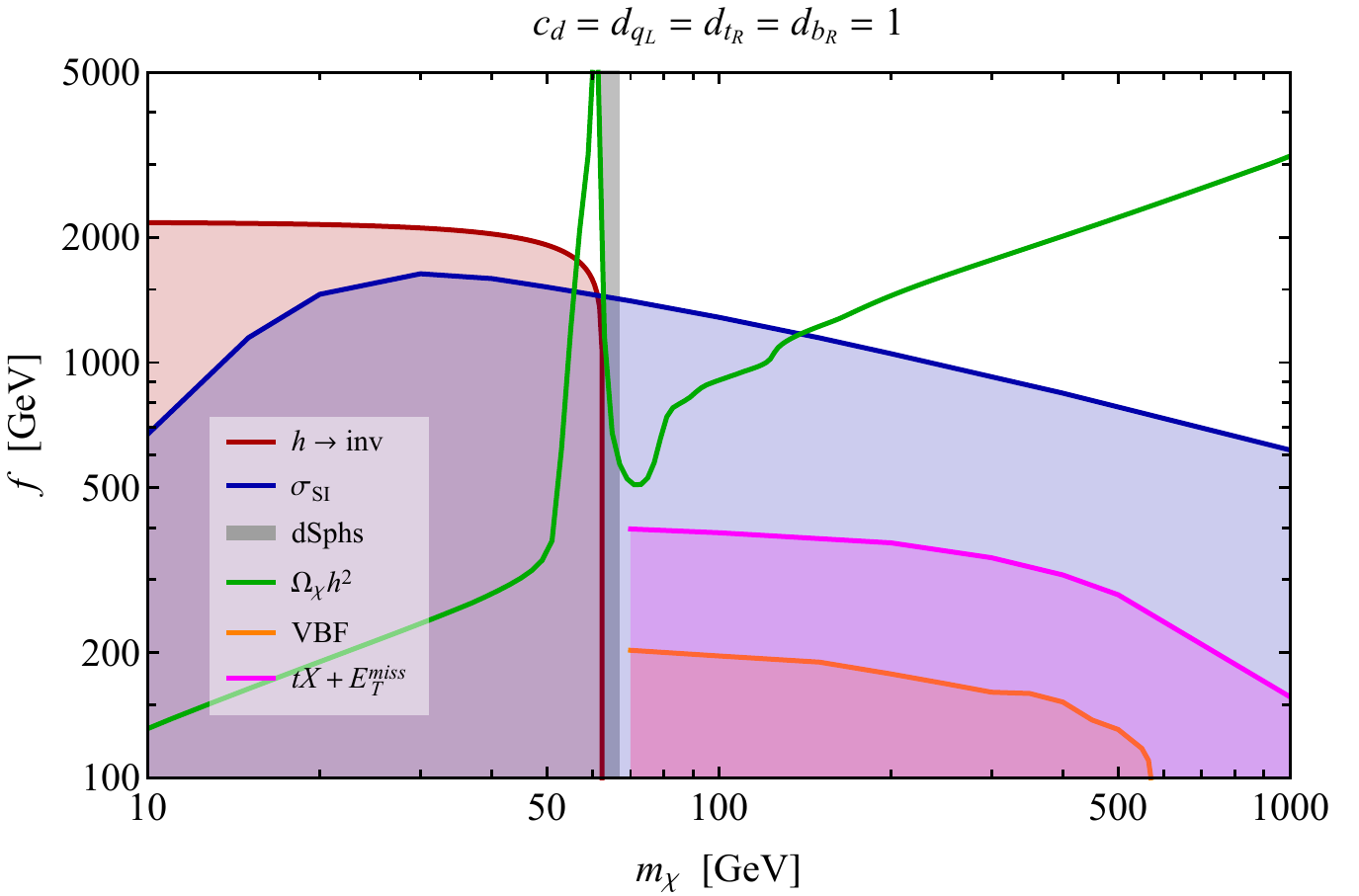} 
\vspace{2mm} 
\caption{\label{fig:summaryS3} As Figure~\ref{fig:summaryS2} but for the  pNGB~DM  benchmark model with $c_d=d_{q_L}=d_{t_R}=d_{b_R}=1$.}
\end{center}
\end{figure}

\begin{figure}[!t]
\begin{center}
\includegraphics[width=0.8\textwidth]{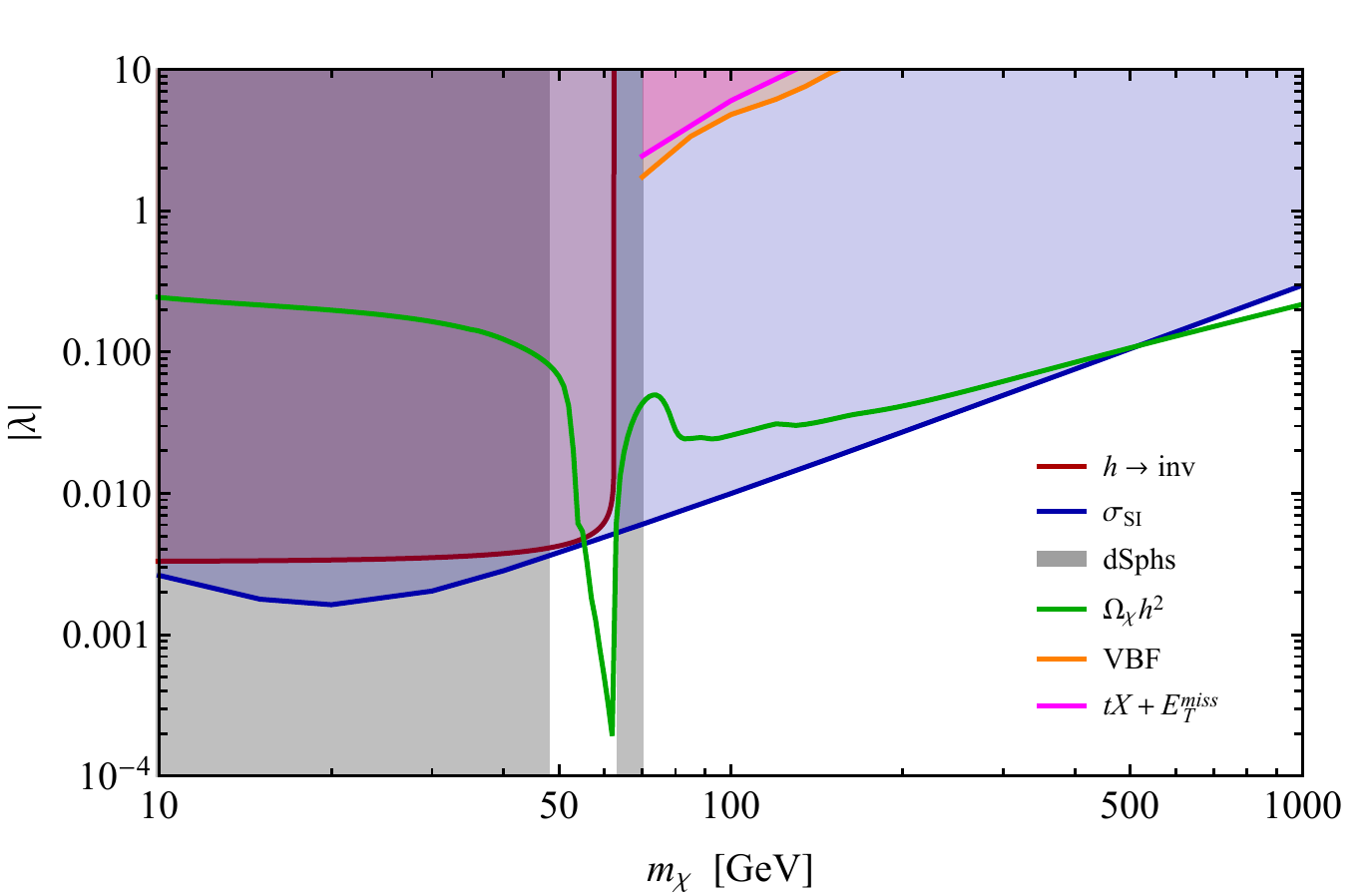} 
\vspace{2mm} 
\caption{\label{fig:summaryS4}  Constraints in the $m_\chi \hspace{0.25mm}$--$\hspace{0.25mm} |\lambda|$~plane for the  marginal Higgs portal model. Apart from the fact that in the parameter space below the green curve the Universe is overclosed, the meaning and colour coding of the shown constraints resemble those of Figure~\ref{fig:summaryS2}. }
\end{center}
\end{figure}

In Figures~\ref{fig:summaryS1} to~\ref{fig:summaryS3}  we summarise the most important constraints in the $m_\chi \hspace{0.25mm}$--$\hspace{0.25mm} f$~plane for the three benchmark models with $c_d=1$, $c_d=c_t=1$ and $c_d=d_{q_L}=d_{t_R}=d_{b_R}=1$. Similar benchmark models have also been considered in~\cite{Ruhdorfer:2019utl}. The pNGB~DM effective field theory  parameters not shown in the headline of each figure are set to zero to obtain the displayed results.  The~dark red and blue regions are excluded by the projected HL-LHC limit on the Higgs invisible branching ratio of ${\rm BR} \left (h \to {\rm inv} \right) < 2.5 \cdot 10^{-2}$~\cite{Cepeda:2019klc} and by the 90\%~CL bounds on the SI DM-nucleon cross section set  by XENON1T~\cite{Aprile:2018dbl}, respectively. The vertical grey bands indicate the DM  mass ranges that are excluded at 95\%~CL by  the $\gamma$-ray observations of dwarf spheroidal galaxies~(dSphs) of the  {\rm Fermi-LAT} and {\rm DES} collaborations in~\cite{Fermi-LAT:2016uux}.~The used experimental bounds assume DM annihilation into~$b \bar b$ final states and that  the measured relic density  is reproduced. The constraints that follow from the latest {\rm Fermi-LAT} search for $\gamma$-ray lines~\cite{Ackermann:2015lka} lead  to weaker constraints on the DM mass of $62.5\,  {\rm GeV} \lesssim m_\chi \lesssim 64 \, {\rm GeV}$ compared to $\chi^\ast \chi \to b \bar b$ even if a favourable  DM distribution $\big($such as an adiabatically contracted Navarro-Frenk-White profile~\cite{Navarro:1995iw}$\big)$ is used to calculate the limits. These bounds are hence not shown in Figures~\ref{fig:summaryS1} to~\ref{fig:summaryS4}. The green curves correspond to the  {\rm PLANCK} value $\Omega_\chi h^2 = 0.12$~\cite{Aghanim:2018eyx}  of the DM relic abundance.  The orange regions displayed in the figures  correspond to the 95\%~CL exclusion limits found in~\cite{Ruhdorfer:2019utl} from a  HL-LHC study of off-shell invisible Higgs production in the~VBF channel.  The magenta domains finally correspond to the 95\%~CL constraints obtained by the  $t X + \etmiss$ analysis strategy discussed in Section~\ref{sec:tXetmissanalysis}. 

In the case of the derivative Higgs portal model, one observes from Figure~\ref{fig:summaryS1} that in the Higgs on-shell region corresponding to $m_\chi < m_h/2$, HL-LHC measurements of invisible Higgs decays  exclude large parts of the parameter space that leads to the correct DM relic density via standard thermal freeze-out. Only a narrow corridor around the Higgs resonance survives this constraint, which is however excluded by DM indirect detection measurements. Since the DM-nucleon scattering rate is momentum suppressed, the stringent limits from   DM direct detection experiments do not put constraints on the pNGB~DM benchmark model with only $c_d = 1$. This opens up the possibility to test such models with $m_\chi > m_h/2$ using mono-$X$ searches at the~HL-LHC, however only if these models lead to a DM underabundance,~i.e.~$\Omega_\chi h^2 < 0.12$. Given that  the VBF limits taken from~\cite{Ruhdorfer:2019utl}  are around $30\%$ better than the $tX+\etmiss$ bounds on $f$, the best test of the derivative Higgs portal model in the Higgs off-shell region seems to be provided by invisible Higgs production in the~VBF channel. In this context it is however important to realise that the study~\cite{Ruhdorfer:2019utl} assumes a  systematic uncertainty on the relevant SM background of $1\%$, while the  shown  $tX+\etmiss$ exclusion is based on  a  systematic uncertainty on the relevant SM background of~$15\%$ (see~Section~\ref{sec:tXetmissanalysis}). Assuming a reduction of the systematic background uncertainties in  $tX+\etmiss$ down to~$5\%$ would bring the VBF and $tX+\etmiss$ exclusion limits closer together. See Appendix~\ref{app:morenumerics} for details. 

As can be seen from Figures~\ref{fig:summaryS2} and~\ref{fig:summaryS3}, the HL-LHC potential to test viable models through mono-$X$ searches is less favourable in the case of the pNGB~DM benchmarks with $c_d=c_t=1$ or $c_d=d_{q_L}=d_{t_R}=d_{b_R}=1$ since in these cases the limits from DM direct detection, though loop suppressed,  turn out to be still severe. In the first case the LL corrections to $\lambda$  in~(\ref{eq:lambdaRGE}) and the finite matching correction to $d_G$ in~(\ref{eq:cBcG}) are both relevant, while in the second case the  LL corrections to~$c_A$  in~(\ref{eq:cBcG}) play an essential role in determining the correct DM direct detection limits. The above LL corrections have not been discussed in the work~\cite{Ruhdorfer:2019utl}, but it is known~(see~for example~\cite{Hill:2011be,Frandsen:2012db,Haisch:2013uaa,Crivellin:2014qxa,Crivellin:2014gpa,DEramo:2014nmf,DEramo:2016gos,Bishara:2018vix}) that the inclusion of radiative corrections can have important effects in the calculation of $\sigma_{\rm SI}$. Comparing the VBF and $tX + \etmiss$ constraints, one sees that in both cases  $c_d=c_t=1$ and $c_d=d_{q_L}=d_{t_R}=d_{b_R}=1$ the limits on $f$ derived here are stronger than the bounds that have been obtained in~\cite{Ruhdorfer:2019utl}. This result follows straightforwardly from the fact that invisible VBF Higgs off-shell production is only sensitive to $c_d$, while the   $tX + \etmiss$ signature receives contributions from  $c_d$ but also from $c_t$, $d_{q_L}$ and $d_{t_R}$. 

In Figure~\ref{fig:summaryS4} we finally summarise the constraints on the marginal Higgs portal model set by DM (in)direct detection experiments, the relic density and future HL-LHC searches.  One observes that  the constraints on $|\lambda|$ from DM direct detection and the HL-LHC are comparable for DM masses $m_\chi < m_h/2$. However, in the case  $m_\chi > m_h/2$ the bounds that follow from~$\sigma_{\rm SI}$ are by more than two orders of magnitude stronger than those that one can hope to obtain at the HL-LHC from mono-$X$ searches. Like in the case of the derivative Higgs portal model,  off-shell invisible Higgs production in the~VBF channel~\cite{Ruhdorfer:2019utl} again seems to be the best way to probe the marginal~Higgs portal model at the LHC if $m_\chi > m_h/2$. This conclusion once more depends on the actual size of systematic background uncertainties of the VBF and $tX+\etmiss$ channels in the HL-LHC environment. Combining the two  mono-$X$ channels as done in the case of  the LHC searches for the invisible Higgs boson decays~(see for instance \cite{ATLAS-CONF-2020-052,ATLAS:2019cid,CMS:2018yfx,CMS:2019bke}) can be expected to improve the ultimate HL-LHC reach.  Performing an actual combination of the VBF and $tX+\etmiss$ channels is however beyond the scope of this article. We add that the potential of the high-energy option of the LHC, the   future circular hadron-hadron collider,  the compact linear collider and  a muon collider  in constraining the marginal Higgs portal through VBF off-shell Higgs production has been studied in the article~\cite{Ruhdorfer:2019utl}. See also~\cite{Matsumoto:2010bh,Kanemura:2011nm,Chacko:2013lna,Craig:2014lda,Ko:2016xwd} for similar analyses.

pNGB~DM models in which both the SM Higgs boson as well as the DM candidate  are composites of  a TeV-scale strongly-coupled sector provide a simultaneous explanation of  the EW hierarchy problem  and the DM puzzle. Key features in this class of beyond the SM theories are that  the SM~Higgs boson and the DM particle are both naturally light, and that  the  leading coupling between  DM and the SM  is the derivative Higgs portal. This portal is strongly suppressed  in the regime of small momentum transfer that is probed by  DM scattering with heavy nuclei, making this type of WIMP easily compatible with the existing strong constraints  from DM direct detection experiments. At~the same time, the interaction strength of DM annihilation turns out to be in the right range to obtain the observed relic density through thermal freeze-out without tuning. However, as we have shown in our work, this simple and attractive picture can be significantly altered by explicit symmetry breaking effects that lead to pNGB~DM interactions beyond the derivative Higgs portal. In fact, once radiative effects are taken into account, only pNGB~DM realisations of the form~(\ref{eq:LEFT}) with~$c_d \neq 0$ and all other pNGB~DM effective field theory  parameters sufficiently small  typically survive the constraints from DM direct detection experiments.  In such scenarios, collider searches for DM production are the only known direct way to explore the pNGB~DM parameter space. If~the DM candidate is kinematically accessible, searches for invisible Higgs boson decays play a key role in such explorations, while  DM masses above the Higgs threshold can be probed by studying mono-$X$ signatures. In~our article, we have extended the earlier study of off-shell invisible Higgs production via~VBF~\cite{Ruhdorfer:2019utl} by developing a search strategy that allows to probe pNGB~DM using~$tX+\etmiss$ signatures. The~$tX+\etmiss$ channels are complementary to VBF Higgs production since they are able to test pNGB~DM interactions like the Yukawa-type DM-top coupling and the current-current type interactions in~(\ref{eq:LEFT})  that are not accessible via the latter mode. Together with~\cite{Ruhdorfer:2019utl} the work presented here provides the blueprints  to search for pNGB~DM at the LHC, and we encourage ATLAS and CMS to perform dedicated experimental searches and interpretations of the relevant mono-$X$ signatures. 

\acknowledgments We thank Maximilian~Ruhdorfer, Ennio~Salvioni and Andreas~Weiler for useful discussions, for their helpful comments on this manuscript and for providing us with  the computer codes employed  in their paper~\cite{Ruhdorfer:2019utl} to determine the DM indirect detection and relic density  constraints on composite pNGB~DM models. Our analytic calculations made use of the {\tt Mathematica} packages {\tt FeynArts}~\cite{Hahn:2000kx}, {\tt FormCalc}~\cite{Hahn:1998yk,Hahn:2016ebn}~and {\tt FeynCalc}~\cite{Mertig:1990an,Shtabovenko:2016sxi,Shtabovenko:2020gxv}. This research has been partially supported by the Collaborative Research Center SFB1258.

\begin{appendix}

\section{Supplementary material}
\label{app:morenumerics}

\begin{figure}[!t]
\begin{center}
\includegraphics[width=0.7\textwidth]{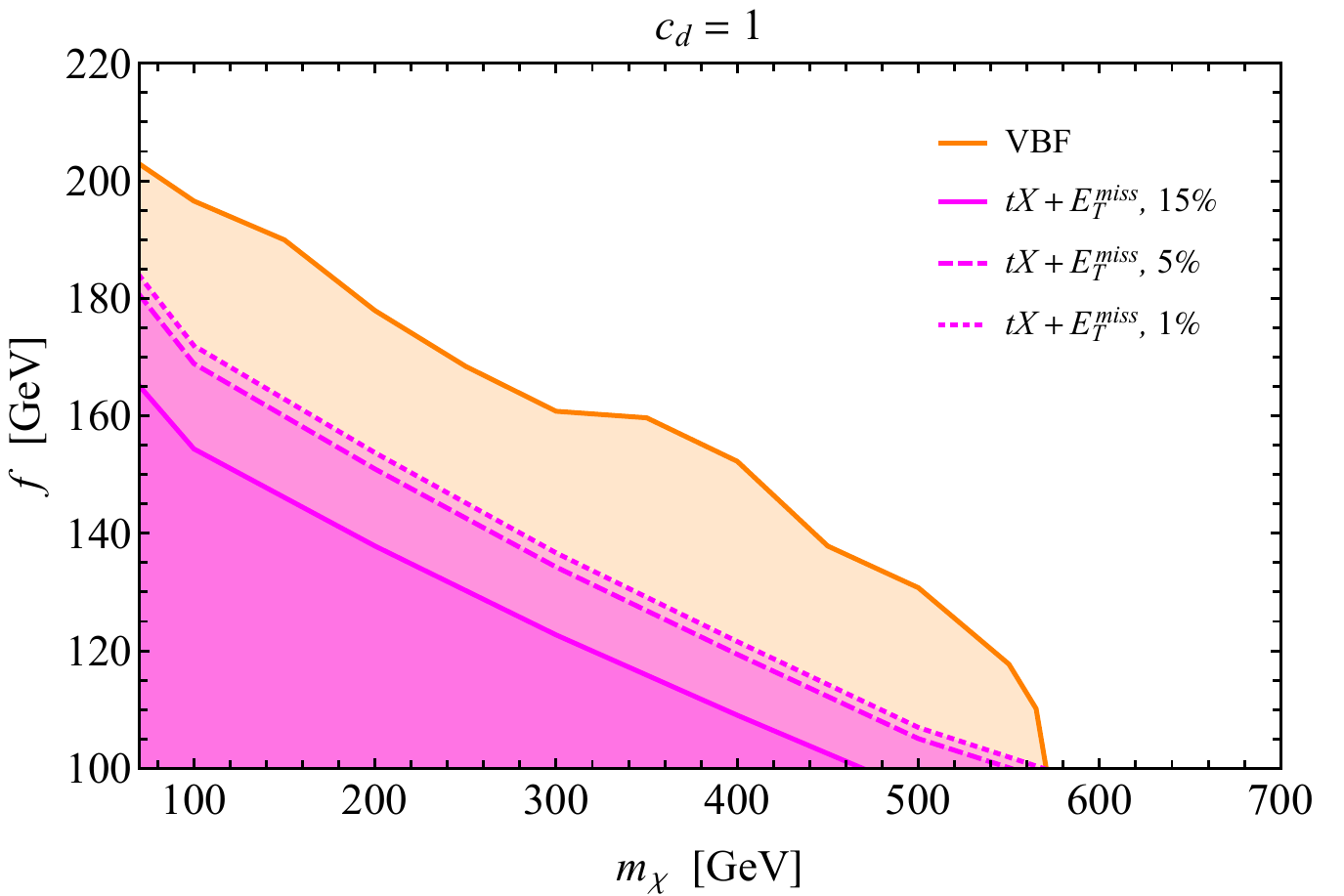} 

\vspace{4mm}

\includegraphics[width=0.7\textwidth]{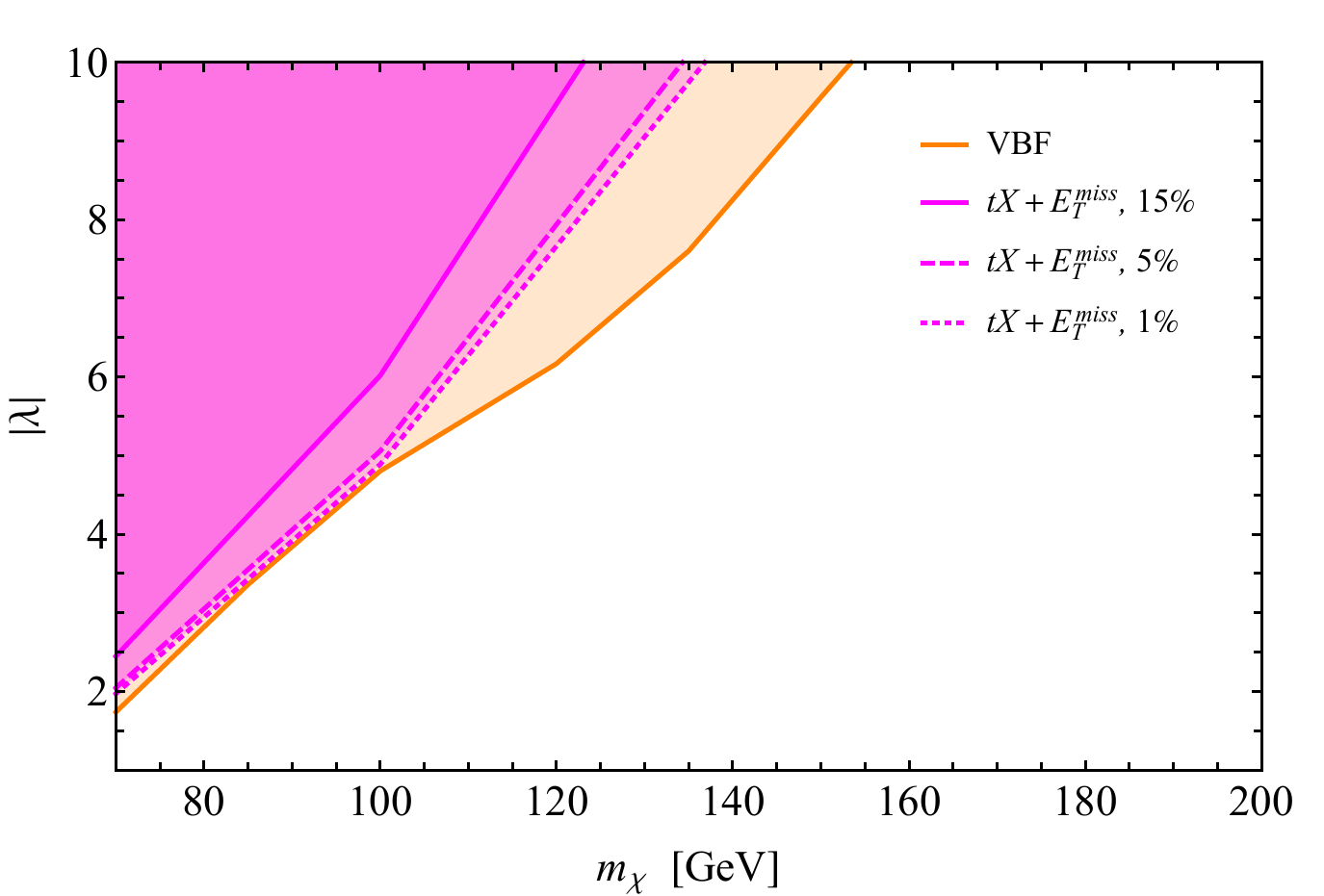} 
\vspace{2mm} 
\caption{\label{fig:supplementary} 95\%~CL constraints in the $m_\chi \hspace{0.25mm}$--$\hspace{0.25mm} f$~plane for the derivative Higgs portal model~(upper panel) and in the $m_\chi \hspace{0.25mm}$--$\hspace{0.25mm} |\lambda|$~plane for the  marginal Higgs portal model~(lower panel). The orange regions  correspond to the 95\%~CL exclusion limits determined in~\cite{Ruhdorfer:2019utl} from a  HL-LHC study of off-shell invisible Higgs production in the~VBF channel, while the magenta contours represent the results of our $t X+\etmiss$ search  assuming a systematic background uncertainty of 15\%~(solid curves), 5\%~(dashed curves) and 1\%~(dotted curves).}
\end{center}
\end{figure}

In this appendix we present HL-LHC projections based on alternative more aggressive assumptions about the systematic uncertainties of our  $tX+\etmiss$ search strategy. Anticipating improvements in detector performance and modelling of SM background processes, we assume that the systematic uncertainties on the number of expected events in the signal regions SR1, SR2 and SR3 are reduced from 15\% to 5\% and 1\%. In Figure~\ref{fig:supplementary}  we show the 95\%~CL constraints in the $m_\chi \hspace{0.25mm}$--$\hspace{0.25mm} f$~plane for the derivative Higgs portal model~(upper panel) and in the $m_\chi \hspace{0.25mm}$--$\hspace{0.25mm} |\lambda|$ plane for the  marginal Higgs portal model~(lower panel).  The orange regions indicate the exclusion limits derived in the study of off-shell invisible Higgs production in the~VBF channel~\cite{Ruhdorfer:2019utl}. The displayed results assume a~1\% systematic uncertainty on the relevant SM backgrounds. For comparison we show in magenta the 95\%~CL limits that derive from the $tX+\etmiss$ search strategy discussed in Section~\ref{sec:tXetmissanalysis}. Here the solid, dashed and dotted contours correspond to assumed systematic background uncertainties of 15\%, 5\% and~1\%, respectively. It is evident from both panels that reducing the systematic uncertainties from 15\% to 5\% has a visible impact on the obtained $tX+\etmiss$ exclusion limits, while a further uncertainty reduction to 1\% has only a minor effect on the bounds in the shown parameter planes. Notice that a reduction of the systematic uncertainties to 5\% may be possible given the steady  progress of both experiment and theory. In the case of  the marginal Higgs portal, such an improvement would lead to a reach in the $tX+\etmiss$ channel that is very similar to the one of VBF invisible Higgs production in the off-shell region. 

\end{appendix}

%\bibliographystyle{apsrev4-1}
%\bibliography{ttMET_pNGB~DM}


\end{document}